\documentclass[fleqn,usenatbib]{mnras}

\usepackage{newtxtext,newtxmath}

\usepackage[T1]{fontenc}
\usepackage{ae,aecompl}

\hypersetup{
colorlinks,
}

\usepackage{graphicx}    
\usepackage{amsmath}    
\usepackage{amssymb}    
\usepackage{gensymb}
\usepackage{multirow}
\usepackage{placeins}   
\usepackage[super]{nth}

\usepackage{color, colortbl}            
\usepackage{xargs}                      
\usepackage[pdftex,dvipsnames]{xcolor}  
\usepackage[draft,colorinlistoftodos,prependcaption,textsize=small]{todonotes}   
\newcommandx{\unsure}[2][1=]{\todo[linecolor=red,backgroundcolor=red!25,bordercolor=red,#1]{#2}}
\newcommandx{\change}[2][1=]{\todo[linecolor=blue,backgroundcolor=blue!25,bordercolor=blue,#1]{#2}}
\newcommandx{\info}[2][1=]{\todo[linecolor=OliveGreen,backgroundcolor=OliveGreen!25,bordercolor=OliveGreen,#1]{#2}}
\newcommandx{\improvement}[2][1=]{\todo[linecolor=Plum,backgroundcolor=Plum!25,bordercolor=Plum,#1]{#2}}

\usepackage{changes}
\colorlet{Changes@Color}{red} 

\usepackage{tikz}
\usetikzlibrary{shapes.geometric, arrows}

\newcommand{\unsim}{\mathord{\sim}} 

\newcommand{\sqd}{\,deg$^{2}$} 
\newcommand{\sd}{\,deg$^{2}$} 
\newcommand{\sdd}{\,deg$^{2}$\,d} 
\newcommand{\psdd}{\,deg$^{-2}$\,d$^{-1}$} 
\newcommand{\fots}{fast optical transients} 

\title[PTF-Sky2Night]{The Palomar Transient Factory Sky2Night programme}

\author[J. van Roestel]
{J. van Roestel$^1$\thanks{j.vanroestel@astro.ru.nl},
P.J. Groot$^1$, T. Kupfer$^{2,3,4}$, K. Verbeek$^1$, S. van Velzen$^5$, M. Bours$^6$, 
\newauthor P. Nugent$^7$, T. Prince$^4$, D. Levitan $^4$, S. Nissanke$^1$, S.R. Kulkarni$^4$, R.R. Laher$^8$ \\
$^1$ Department of Astrophysics/IMAPP, Radboud University Nijmegen, P.O.Box 9010, 6500 GL, Nijmegen, The Netherlands \\
$^2$ Kavli Institute for Theoretical Physics, University of California, Santa Barbara, CA 93106, USA\\
$^3$ Department of Physics, University of California, Santa Barbara, CA 93106, USA\\
$^4$ Cahill  Center  for  Astronomy  and  Astrophysics,  California
Institute of Technology, Pasadena, CA 91125 \\
$^5$ Center of Cosmology and Particle Physics, New York University, New York, NY 10003, USA \\
$^6$ Instituto de F\'{i}sica y Astronom\'{i}a, Universidad de Valpara\'{i}so, Avenida Gran Bretana 1111, Valpara\'{i}so 2360102, Chile \\
$^7$ Lawrence Berkeley National Laboratory, UC Berkley Department of Astronomy, Berkeley, CA, USA \\
$^8$ Spitzer Science Center, California Institute of Technology, Pasadena, CA 91125, USA}

\date{Accepted XXX. Received YYY; in original form ZZZ}

\pubyear{2018}

\begin{document}
\label{firstpage}
\pagerange{\pageref{firstpage}--\pageref{lastpage}}
\maketitle

\begin{abstract}
We present results of the Sky2Night project: a systematic, unbiased search for \fots\ with the Palomar Transient Factory. We have observed 407\,\sd\ in $R$-band for 8 nights at a cadence of 2 hours. During the entire duration of the project, the 4.2\,m William Herschel Telescope on La Palma was dedicated to obtaining identification spectra for the detected transients. During the search, we found 12 supernovae, 10 outbursting cataclysmic variables, 9 flaring M-stars, 3 flaring active Galactic nuclei and no extragalactic \fots. Using this systematic survey for transients, we have calculated robust observed rates for the detected types of transients, and upper limits of the rate of extragalactic \fots\ of $\mathcal{R}<37\times 10^{-4}$\,\psdd\ and $\mathcal{R}<9.3\times 10^{-4}$\,\psdd\ for timescales of 4\,h and 1\,d and a limiting magnitude of $R\approx19.7$. We use the results of this project to determine what kind of and how many astrophysical false positives we can expect when following up gravitational wave detections in search for kilonovae.
\end{abstract}



\section{Introduction}
\begin{figure*}
  \centering
  \includegraphics[width=0.99\textwidth]{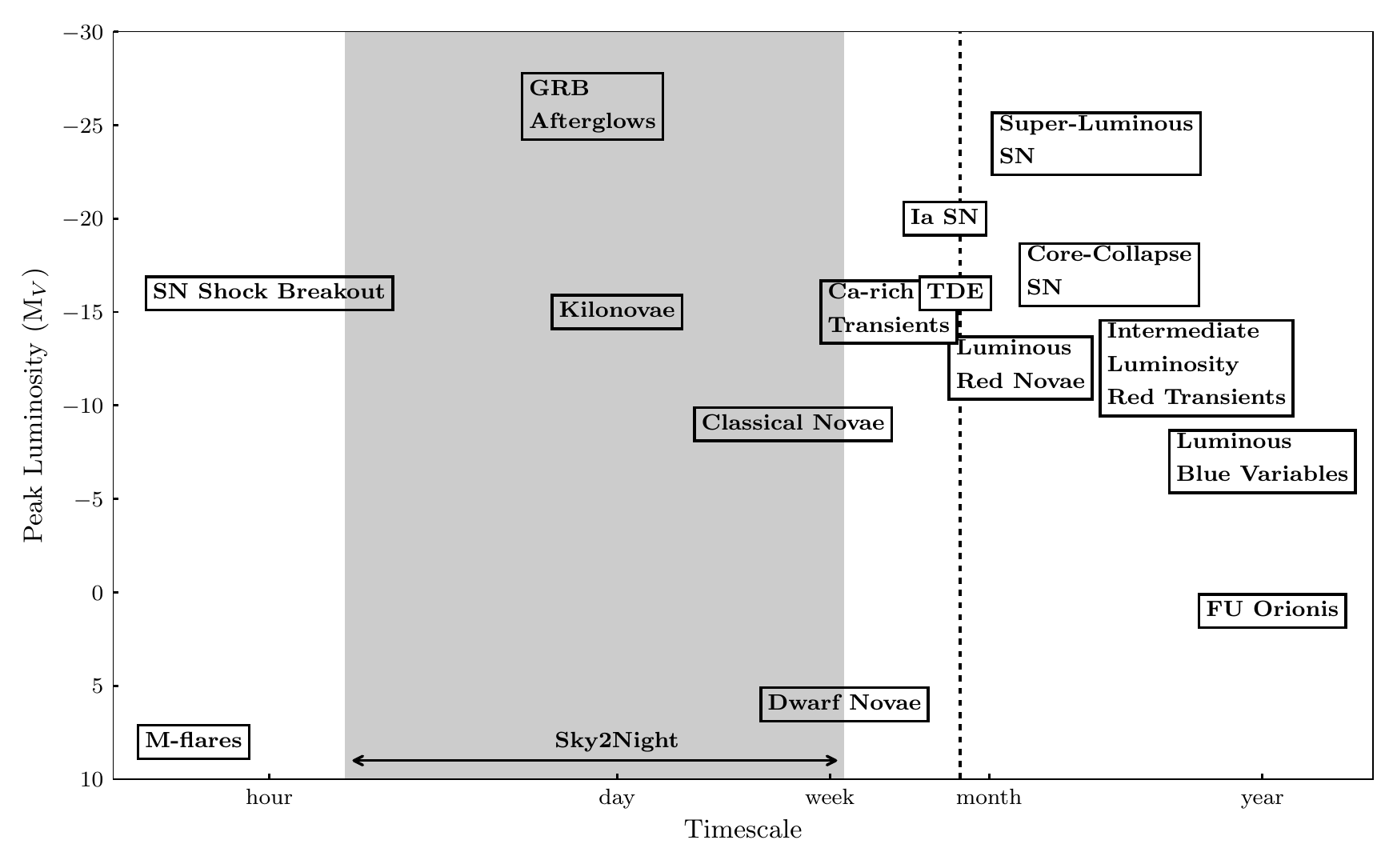} \label{fig:transientphasespace}
  \caption{The absolute luminosity versus the typical timescale of transients. The grey area shows the timescales which are probed by the Sky2Night project. The dashed line shows the time between the acquisition of the reference images and the end of the survey. This figure is adapted from \citet{2011PhDT........35K}.}
  \label{fig:phasespace}
\end{figure*}

Fast optical transients are transients which appear and disappear within 24 hours or less. The rate of fast optical transients is not well known (see Fig. \ref{fig:transientphasespace}). The reason why the fast transient sky has not yet been systematically explored is due to technical limitations. To find fast transients a high cadence is required, which means that area and/or depth need to be sacrificed. For example, a 3-day cadence supernova survey can cover an area 100 times larger than a survey of optical transients with a cadence of 1 hour (using the same setup). In addition, follow up of \fots\ is difficult since it requires rapid detection and identification of the transient and triggering of a follow-up telescope.

For this reason, almost all known extragalactic \fots\ (timescales of less than 1 day) 
have been found as a counterpart of a transient detected at another wavelength where larger solid angle facilities are possible, e.g. X-ray or gamma-ray satellites. The most well studied are gamma-ray burst afterglows: interactions between highly relativistic outflows (jets) and their environment \citep[e.g.][]{1999PhR...314..575P}. Although they can be bright, because of the low rate \citep[$\approx 1000\,\mathrm{yr^{-1}}$ with $R<20$,][]{2015ApJ...803L..24C} and in particular because of their rapid fading \citep[$\unsim$ magnitudes per hour, e.g.][]{2015ApJ...806...52S,2015ApJ...815..102F} they are very difficult to find in blind searches. So far, only one\footnote{a strong candidate without a gamma-ray detection is described by \citet{2013ApJ...769..130C}} GRB afterglow has been found in a blind search: iPTF14yb \citep{2015ApJ...803L..24C}.

In searches for fast extragalactic transients, many Galactic \fots\ are detected: outbursts of stars in our Galaxy with short ($\unsim 1$\,d) timescales.  
These are sometimes considered a `foreground fog', but they are also interesting to study in their own right. Flaring M-dwarfs are the most common Galactic \fots\ with typical time scales of 10-100 minutes \citep{2014ApJ...797..121H,2016ApJ...829..129S} up to 7 hours \citep{2010ApJ...714L..98K} and outburst magnitudes up to 9 magnitudes in V-band \citep[e.g.][]{2014ApJ...781L..24S}. Understanding M-flare rates and intensities have recently become important with regards to planetary habitability \citep[e.g.][]{2017ApJ...841..124V}. The other type of common Galactic transients are eruptions in compact binary systems with an accretion disk. The most common are dwarf novae (DN), caused by accretion disc instabilities in systems where a white dwarf accretes mass from a main sequence companion. These outbursts can brighten the system by up to 8 magnitudes \citep[e.g. WZ Sge,][]{2004AJ....127..460H}, with short rise timescales ($\unsim$ 1 day) and can last for a few days to weeks \citep{2003cvs..book.....W}. 

In 2017, the aLIGO/aVirgo gravitational wave observatories \citep{2015CQGra..32g4001L,2015CQGra..32b4001A} detected the first binary neutron star (BNS) merger \citep{PhysRevLett.119.161101}.
Rapid optical follow up of this event resulted in the discovery of the optical counterpart  AT~2017gfo (\citealt{2017ApJ...848L..12A,2017ApJ...848L..13A}, see also: \citealt{
2017PASA...34...69A,
2017Natur.551...64A,
2017Sci...358.1556C,
2017ApJ...848L..17C,
2017ApJ...848L..29D,
2017Sci...358.1570D,
2017Sci...358.1565E,
2017SciBu..62.1433H,
2017Sci...358.1559K,
2018NewA...63...48L,
2017Natur.551...67P,
2018ApJ...852L..30P,
2017Sci...358.1574S,
2017Natur.551...75S,
2017ApJ...848L..27T,
2017Natur.551...71T,
2017PASJ...69..101U,
2017ApJ...848L..24V}).
The optical counterpart, called a kilonova, had been theorized to accompany a BNS merger by \citet{
1998ApJ...507L..59L, 
2005astro.ph.10256K, 
2010MNRAS.402.2771M, 
2011ApJ...736L..21R, 
2013ApJ...775...18B, 
2013ApJ...775..113T, 
2014MNRAS.441.3444M, 
2015MNRAS.450.1777K}.  
AT~2017gfo is consistent with the kilonova model predictions: it was fading rapidly ($\Delta r\approx\,1.2\,\mathrm{mag\,d^{-1}}$), had a peak absolute magnitude of $M_r\approx-16\,$mag and displayed rapid reddening ($\approx\,0.8\,\mathrm{mag\,d^{-1}}$ in $g-r$). 

The optical signal can be well modelled using either two or three outflow components. The two-component models are a combination of a rapidly fading `blue' component, emitted by fast-moving, low opacity material, and a slower fading `red' component emitted by slower-moving, high opacity material. Three component models add an addition `purple' component, with intermediate velocity and opacity. \citet{2017ApJ...851L..21V} show that a three-component model is the best explanation for kilonova AT~2017gfo. However, we should be aware that future kilonovae can be quite different than AT~2017gfo. For example, a lower amount of ejected mass and a different viewing angle results in a kilonovae which is significantly fainter, from $M_R=-16$ down to $M_R=-12$ \citep[e.g.][]{2015MNRAS.450.1777K,2017CQGra..34j4001R}.

Currently, the aLIGO/aVirgo detectors are being upgraded, increasing the distance (and thus volume) at which they can detect BNS mergers. However, the localisation of the events will remain relatively poor \citep[120-180\,\sqd, depending on the SNR of the event, ][]{2016LRR....19....1A}. This means after the detection of a BNS merger, optical telescopes will still have to search a large area to find the faint optical counterpart, $R\approx 19.5$\,mag at 200\,Mpc if it is similar to AT~2017gfo. One of the problems is that in such a large area and magnitude limit, many other (fast) transients will appear which can be confused for a kilonova; a `needle-in-the-haystack' problem.

In the last decade, there have been a few studies that performed a blind search for fast transients in an attempt to determine the observed rate of \fots. One of the earliest attempts was by \citet{2004ApJ...611..418B}. They carried out an unbiased transients search on the data from the Deep Lensing Survey \citep[DLS,][]{2002SPIE.4836...73W} and found two M-flares and one potential extragalactic transient (OT20030305). However, follow up of the quiescent counterpart by \citet{2006ApJ...644L..63K} shows that OT20030305 is also a flaring M-star. Other searches also found only Galactic transients, mainly cataclysmic variables (CVs) and M-dwarf flares. For example, \citet{2005ApJ...631.1032R} used the ROTSE-II survey to search for untriggered GRBs but found only six outbursting cataclysmic variables. \citet{2008ApJ...682.1205R} performed a high cadence survey on the Fornax galaxy cluster (cadence 32\,min, depth B=22\,mag). They also did not find any extragalactic \fots\ in their search. The first multi-colour search for \fots\ was performed by \citet{2013ApJ...779...18B}, who showed that colours are very useful in identifying the transients. In their search for fast transients, they only found flares on faint M-stars and slow-moving asteroids.
More recently, \citet{2017arXiv171002144C} and \citet{2018PASJ...70....1U} performed multi-colour surveys with the goal to measure the rate of false positives when searching for kilonovae.

In this paper, we present an 8-day unbiased search for all transients in 407\sqd\ of the sky. The search was combined with rapid, unbiased spectroscopic follow-up. To identify the transients we used the Palomar Transient Factory (PTF) 
and for immediate (within 24 hours) spectroscopic follow-up we used the William Herschel Telescope \citep[WHT,][]{1985VA.....28..531B}. The main goals of the project are to measure the rate of intra-night transients (Galactic and extragalactic) and to determine the expected types of false positives when searching for the optical counterpart of gravitational waves by BNS mergers. The survey design is discussed in Section \ref{sec:surveydesign}. The execution of the observations and data reduction are described in Section \ref{sec:observations}. The results: the survey characteristics, an overview of all detected transients, and the observed transient rates are presented in Section \ref{sec:results}. We discuss the results in Section \ref{sec:discussion}. The last section summarizes the paper and lists the main conclusions.

\section{Survey design}\label{sec:surveydesign}
The project consists of two parts: identification of transients with PTF, and rapid spectroscopic classification with the WHT. 

To search the sky for transients, PTF used the 48-inch Oschin Schmidt Telescope at Palomar Observatory (P48), equipped with the CFH12K camera. The mosaic camera consists of 11 working CCDs with 4k$\times$2k pixels each. The system has a pixel scale of $1.01\arcsec$/pixel and a total field of view of 7.26 square degrees \citep{2009PASP..121.1334R,2009PASP..121.1395L}.
P48 was available for 8 nights of dark time on 2010 November 1--8 (in MJD range 55501.08--55508.85). We used the standard PTF setup of 60\,s exposure times and the $R_\mathrm{Mould}$ filter ($R$ in the rest of the paper).
We chose a target cadence of 2 hours and observed the same 59 PTF fields every night (see Table \ref{tab:S2Nfields} for an overview). The fields are adjacent to each other on the sky and slightly overlapping, so the effective area covered is 407\sqd. The fields were selected such that they were observable the entire night and are located at an intermediate Galactic latitude ($-45\degree<b<-25\degree$), allowing us to study both Galactic and extragalactic transients (see Fig.~\ref{fig:transientphasespace}). The ecliptic latitude of the fields is between $-3 \degree < \lambda < 15\degree$.

To be able to rapidly identify fast transients, PTF used an automated image processing pipeline which does bias and flatfield corrections, source extraction and image subtraction on all new images \citep{2016PASP..128k4502C}. Reference images of the target fields were obtained 14--16 days before the start of the project. Each reference image was constructed using at least 5 individual images. The difference images were presented to human scanners to identify transient candidates of interest and reject false positives. The best candidates were marked for follow-up spectroscopy.

We used the 4.2\,m William Herschel Telescope (WHT) on La Palma, Spain, to obtain classification spectra of new transients. The WHT was dedicated for this purpose for 7 nights starting after the first night of PTF observations (MJD 55501.75). The instrument used to observe the transients was ACAM \citep{2008SPIE.7014E..6XB}. ACAM is both an imager and low-resolution spectrograph ($\mathrm{R}\approx500$, 4000--9000\,\AA) and is therefore ideally suited for rapid transient identification. We first observed the candidate transients with ACAM in imaging mode to confirm if they are real and to determine the brightness, followed by an identification spectrum.

\section{Observations}\label{sec:observations}


\begin{figure}
  \centering
  \includegraphics[width=0.99\columnwidth]{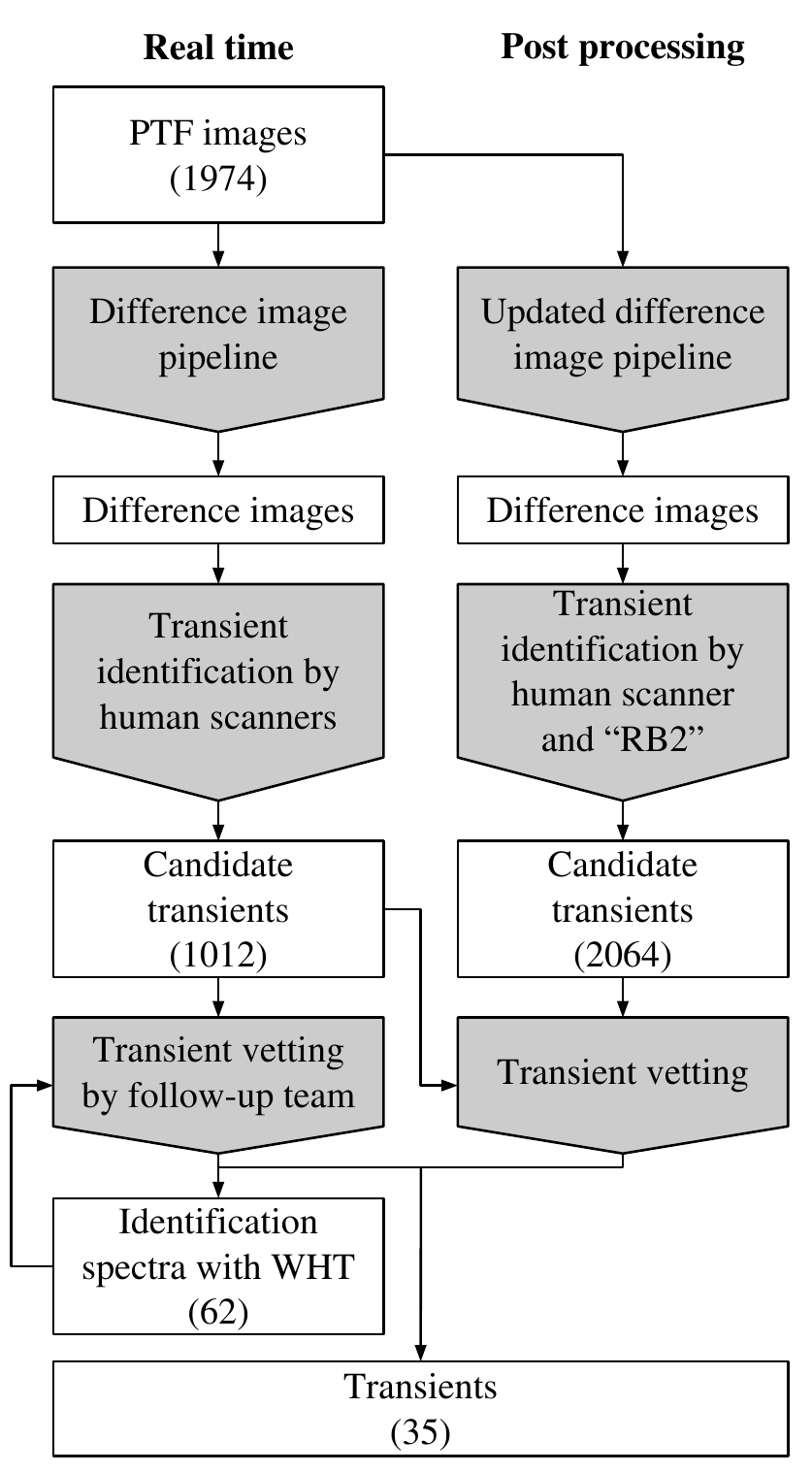}
  \caption{The Sky2Night data analysis and transient detection procedure. The left column shows the real-time analysis of the data, the right column shows the re-analysis of the data. White boxes show data products, within brackets the number of items, grey boxes indicate operations/filters which are applied to the data.}
  \label{fig:dataflow}
\end{figure}

During the project, the weather at the P48 was mostly good. Fifteen per cent of the time was lost due to bad weather, mostly during nights 2, 7, and 8. The seeing was typically 2.5\arcsec, but it was highly variable and regularly increased up to 4\arcsec\ (Table \ref{tab:weather}). A total of 1974 exposures were obtained, with a median of 5 exposures per field per night. Fewer observations were obtained during nights 2, 7, and 8; with a median of 3, 3, and 2 observations per field. A full overview of all PTF observations can be found in Table \ref{tab:nobs}. 

Fig. \ref{fig:dataflow} shows a schematic overview of the data reduction and transient detection process. The new images were processed and difference images were created by the PTF pipeline as quickly as possible (see Sec. \ref{sec:surveydesign}). This ranged from half an hour to a few hours after the observation because the image processing pipeline could not keep up with the flow of incoming images. As soon as the difference images were available they were analysed by multiple human scanners to identify new transients. Identifying the real transients on the difference images was not trivial since the difference images contained many artefacts (e.g. slight misalignments of the images, bad pixels). To reduce the time spend on visual inspection of candidates, we only inspected candidate transients which had two or more detections. This was done to get rid of asteroids but also filtered out some of the artefacts. In addition, the human scanners were assisted by a machine learning algorithm to get rid of the most obvious false positives (a rudimentary version of the `RealBogus' pipeline, see \citealt{2012PASP..124.1175B}, \citealt{2016PASP..128k4502C} and also \citealt{2011MNRAS.412.1309S}). A total of 1012 candidate transients were judged by the human scanners to be potentially real, and these were passed on for further inspection by the follow-up team at the WHT.

\begin{table}
\centering
\caption{The different types of transient candidates found in the real-time search.}
\label{tab:transientcandidates}
\begin{tabular}{ll}
Type & \# \\
\hline
Point-source counterpart &  \\
- Star & 873 \\
- QSO & 15 \\
- Faint star ($R \gtrsim 20.5$) & 3 \\
Galaxy counterpart & \\
- Nuclear & 48 \\
- Near galaxy & 12 \\
Artefact  & 26 \\
Moving object & 35 \\
\hline
Total & 1012 \\
\end{tabular}
\end{table}

The 1012 candidate transients were more carefully vetted by the follow-up team by inspecting the images obtained by PTF and, if available, SDSS images \citep{2009ApJS..182..543A}. In addition, we checked if the target corresponded to a known object in SIMBAD database\footnote{\url{http://simbad.u-strasbg.fr/simbad/}}. An overview of the different kinds of transient candidates is given in Table \ref{tab:transientcandidates}. The bulk of the potential transients were associated with a known point-like source. The majority (873) of these were either due to bad subtraction of a star or low amplitude variability of a star. Besides variable stars, there were also 15 QSOs which brightened significantly during the project. We found three transient candidates without a counterpart in the PTF images, but for which a point-source was detected in the SDSS images. For all three, the SDSS and Pan-STARRS catalogues indicate that they are pointsources. A number of potential transients (60) were found near a galaxy. The majority of these (48) were at the core of the galaxy, and it is difficult to determine if this is a bad subtraction, AGN activity, or a supernova in the core of a galaxy. Five of these could be matched to a known AGN. Experience with other PTF data indicates that the remaining nuclear transients are likely bad subtractions or AGN. The 12 remaining transients with a nearby galaxy were strong supernova candidates.
A few objects (35) were initially flagged as transients but were later identified as moving objects. In addition, 26 candidates were caused by artefacts (bad pixels, very bright stars and `ghosts').

The PTF imaging data was thoroughly re-analysed in 2014 to make sure no transients were missed during the initial search (the right column in Fig. \ref{fig:dataflow}). The images were reprocessed using an improved version of the image processing pipeline. New difference images were made. All sources present on the new difference images were analysed using the `RealBogus2' pipeline \citep{2013MNRAS.435.1047B,2016PASP..128k4502C}. We used a lower than normal threshold value of 0.3, compared to the 0.53 advised in the paper. This lower threshold corresponds to a missed detection rate of 5 per cent (compared to 10 per cent for a threshold of 0.53). This second search resulted in 2064 candidates transients, of which 105 overlap with the initial sample of 1012 sources. All these transient candidates were vetted using PTF, SDSS and Pan-STARRS \citep{2016arXiv161205560C} images and CRTS light curves \citep{2009ApJ...696..870D} and also using the SIMBAD database. This re-analysis recovered all real transients identified by the human scanners during the Sky2Night run. However, we also identified 2 faint supernovae and 5 flaring M-stars in the overlapping sample of transients. In addition, one new flaring M-star was found that was missed entirely during the initial search.

The most promising candidate transients found during the real-time search, typically transients candidates which were bright or were rapidly getting brighter, or were located off-centre from a galaxy, were observed with the WHT telescope.
The results are shown in Table \ref{tab:SNstats}. The majority of the supernovae (8) are of type Ia. Since we obtained both a spectrum and an 8-day light curve, there is little uncertainty in the classification. The subtypes are more difficult to determine, but all except two, appear to be normal type Ia supernovae \citep[for an overview of Ia subtypes, see for example][]{2017arXiv170300528T}. PTF10aaiw, for which we have two spectra, is a ``91T''-like supernova \citep{1992ApJ...384L..15F} according to the cross-correlation with template spectra. The spectra of PTF10aaiw show a shallow Si\,\textsc{ii} 6355\,\AA\ absorption line and deep Fe\,\textsc{iii} absorption features, which are the main features discriminating ``91T''-type supernovae from normal SNIa supernovae. In addition, the absolute peak magnitude as determined from the light curve fit, $M_B\approx -19.5$, is consistent with being a ``91T''-type supernova. PTF10zej could be a ``91bg''-type supernova \citep{1992AJ....104.1543F}; the obtained spectra match almost equally well with ``91bg''-templates and normal Ia-template spectra. The estimated absolute magnitude is only $M_B\approx-18.0$, which puts it at the boundary between normal Ia supernova and ``91bg''-type supernovae. With the available data, we cannot make a certain sub-classification of the subtype of PTF10zej.
The results are shown in Table \ref{tab:SNstats}. The majority of the supernovae (8) are of type Ia. Since we obtained both a spectrum and an 8-day light curve, there is little uncertainty in the classification. The subtypes are more difficult to determine, but all except two, appear to be normal type Ia supernovae \citep[for an overview of Ia subtypes, see for example][]{2017arXiv170300528T}. PTF10aaiw, for which we have two spectra, is a ``91T''-like supernova \citep{1992ApJ...384L..15F} according to the cross-correlation with template spectra. The spectra of PTF10aaiw show a shallow Si\,\textsc{ii} 6355\,\AA\ absorption line and deep Fe\,\textsc{iii} absorption features, which are the main features discriminating ``91T''-type supernovae from normal SNIa supernovae. In addition, the absolute peak magnitude as determined from the light curve fit, $M_B\approx -19.5$, is consistent with being a ``91T''-type supernova. PTF10zej could be a ``91bg''-type supernova \citep{1992AJ....104.1543F}; the obtained spectra match almost equally well with ``91bg''-templates and normal Ia-template spectra. The estimated absolute magnitude is only $M_B\approx-18.0$, which puts it at the boundary between normal Ia supernova and ``91bg''-type supernovae. With the available data, we cannot make a certain sub-classification of the subtype of PTF10zej.
During the spectroscopic follow-up, the weather was variable. Most nights were clear, but during nights 2 and 3 time was lost due to passing clouds. Although night 7 was clear, about half the time was lost due to high humidity. The seeing during the nights varied between 0.8\arcsec and 4\arcsec. Table \ref{tab:weather} shows an overview of the weather conditions at the WHT. A total of 62 transient candidates were observed. Exposure-times of the spectra range between 300 seconds and 1200 seconds. For calibration, either standard star SP2157+261 or SP0804+751 were observed at the beginning or end of the night. A quick reduction of the spectra was performed within 24 hours in order to identify any events which might need additional follow-up. The data were later reduced using standard procedures using \textsc{iraf}. For some transients, spectra were obtained with other telescopes as part of the PTF collaboration and were also used in the identification of transients (see also Table \ref{tab:transientoverview}). All these spectra (including header information) are available on Wiserep\footnote{https://wiserep.weizmann.ac.il/} \citep{2012PASP..124..668Y}.

\section{Results}\label{sec:results}
\subsection{Survey characteristics}
\begin{figure*}
  \centering
  \includegraphics[width=0.99\columnwidth]{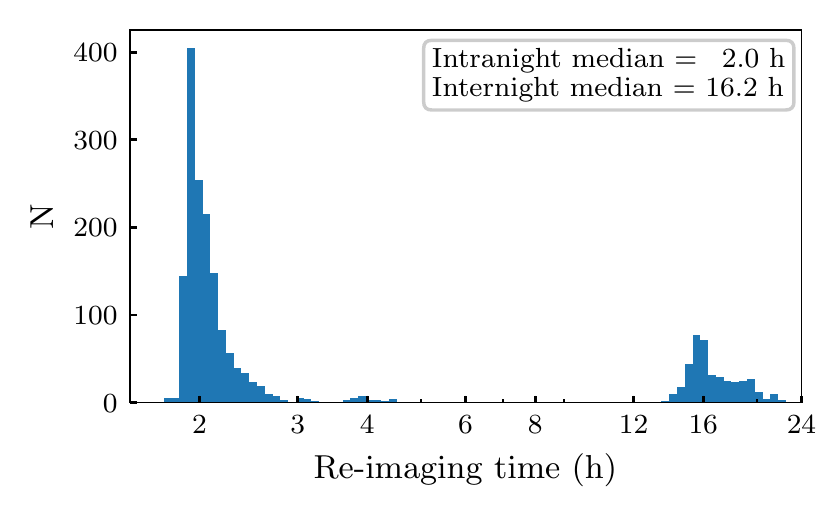}
  \includegraphics[width=0.99\columnwidth]{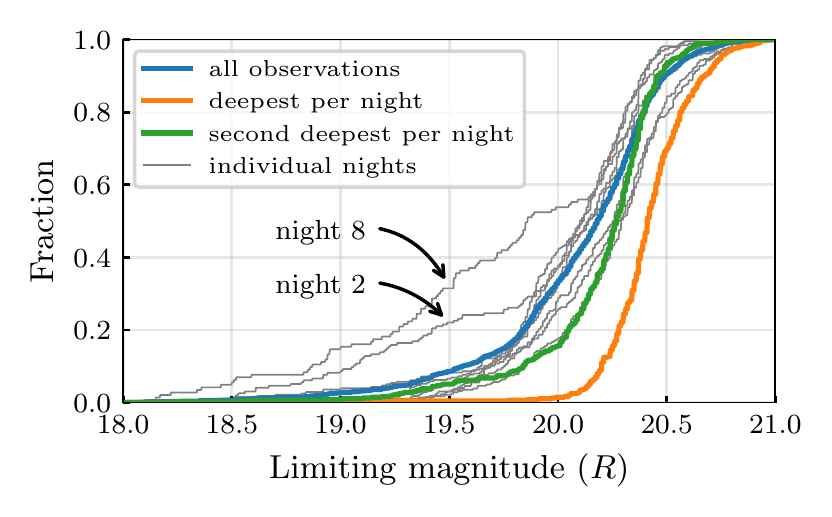}
  \includegraphics[width=0.99\columnwidth]{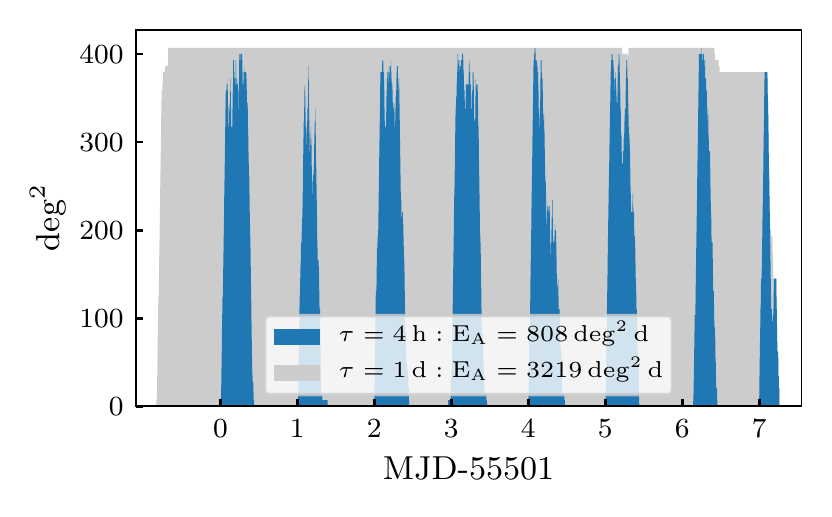}
  \includegraphics[width=0.99\columnwidth]{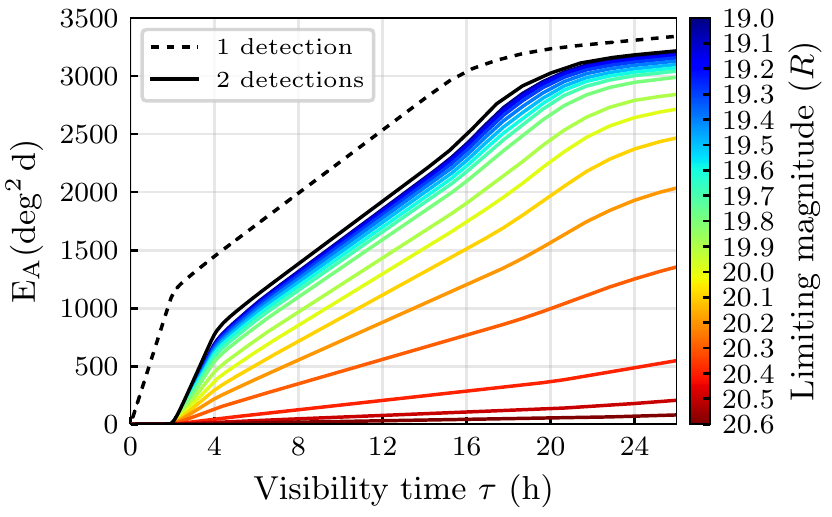}
   \includegraphics[width=0.99\columnwidth]{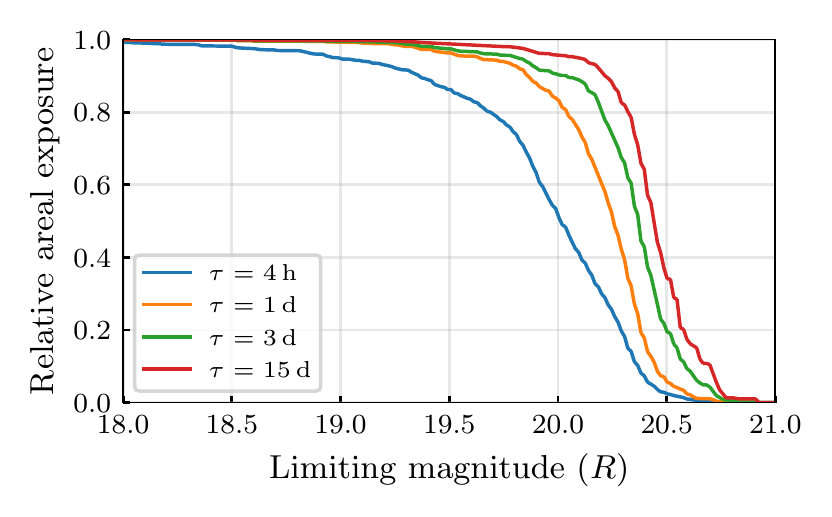}
 \caption{The survey characteristics. 
 (\textit{top-left}) The distribution of time between observations for all fields. 
 (\textit{top-right}) The cumulative histogram of the limiting magnitudes of all observations.
  (\textit{mid-left}) The number of square degrees in which we are able to find transients with 2 detections as a function of time. The area under the curve is the areal exposure ($E_A$).
  (\textit{mid-right}) The areal exposure ($E_A$) as function of visibility time ($\tau$). The black line is the areal exposure if all observations are used, while the coloured lines show the areal exposure if observations that reach a certain limiting magnitude are used.
    (\textit{bottom}) The fraction of the total areal exposure (the black line in the mid-right panel) as function of magnitude for different visibility times.
 }
  \label{fig:survey_characteristics}
\end{figure*}

\begin{figure*}
  \centering
  \includegraphics[width=0.99\textwidth]{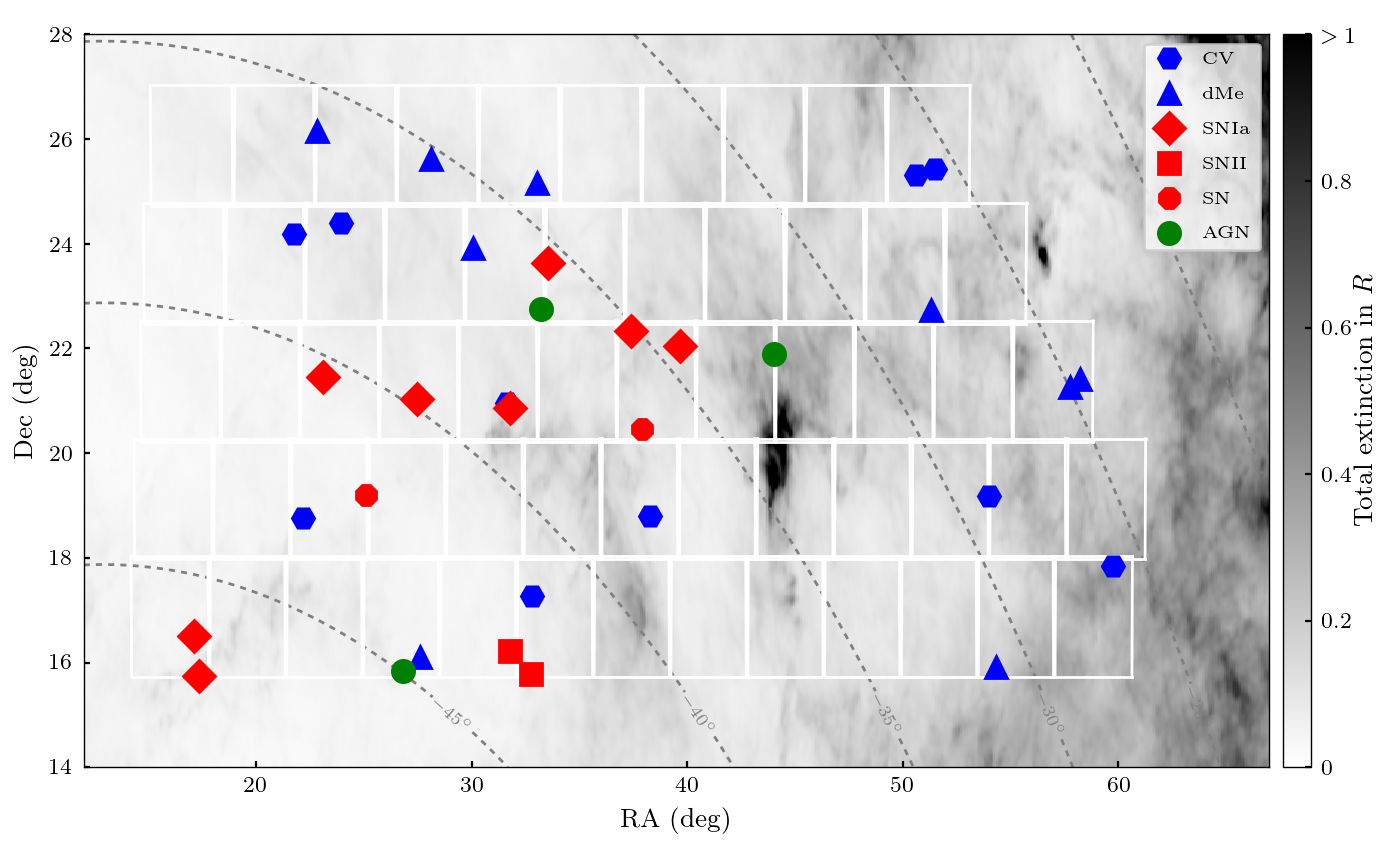}
 \caption{The observed area and all transients found during the project. The white rectangles indicate the 59 PTF fields. The dashed lines indicate different Galactic latitudes. The grey background indicates the total amount of Galactic extinction \citep{1998ApJ...500..525S}.}
  \label{fig:FOV}
\end{figure*}

An overview of the most important metrics of the survey is given in Fig.~\ref{fig:survey_characteristics}. The time between observations, generally referred to as cadence, is given in the top left panel in Fig.~\ref{fig:survey_characteristics}. The median time between observations is 2.0\,h within a night and almost all fields have been revisited within 3 hours. There is also a longer delay of about 16.8\,h between revisits, which is due to the day-night cycle.

The limiting magnitude of the observations are shown in the top right panel. We empirically measure the limiting magnitude by calculating the \nth{95} percentile magnitude of all sources detected in the image. The source detection by PTF is performed with \textsc{Sextractor} \citep{1996A&AS..117..393B} with a detection threshold of 4 standard deviations above the background noise. The median limiting magnitude of all observations is $R=20.18$\,mag, with 95 per cent of the observations in the range of $R=18.92$\,mag and $R=20.70$\,mag (top right panel in Fig.~\ref{fig:survey_characteristics}). The figure also shows the distribution of the limiting magnitude of the deepest image per night (median $R=20.41$\,mag) and the second deepest image per night (median $R=20.27$\,mag), which is the relevant measure for transients which are visible for longer than 1 day. In addition, the distribution for individual nights is plotted, which shows that the nights are comparable, except for nights 2 and 8, due to clouds. Note that the limiting magnitudes are not randomly distributed in each night, but vary as a function of time in the night (see Fig. \ref{fig:PTFlimmag}). This is caused by the airmass related extinction as the fields are tracked from horizon to horizon. At the beginning of the night, the limiting magnitude is about $R \approx 20.0$\,mag per exposure and increases to $R \approx 20.4$\,mag during the middle of the night, and then decreases again to $R \approx 20.0$\,mag toward the end of the night. A few spikes can be seen in the limiting magnitudes as a function of time which are caused by passing clouds.

In order to calculate an observed rate of transients, we need to determine how much area we have effectively monitored and for how long: the areal exposure $E_\mathrm{A}(\tau)$ (units: \sdd), which is a function of how long a transient is visible ($\tau$). To calculate the areal exposure for Sky2Night, we test if a transient (visible for a set duration $\tau$) would have been detected at least twice in our survey. The result for transients visible for $\tau$=4\,h and $\tau$=1\,d is shown as the shaded area in the mid-left panel in Fig.~\ref{fig:survey_characteristics}. The total areal exposure for a given visibility time can be calculated by integrating over time (i.e. the area of the shaded surface in the mid-left panel). The areal exposure as a function of visibility time ($\tau$) is shown in the mid-right panel. Here we also show the areal exposure if we take the limiting magnitude of the images into account. The bottom panel shows the same information in a different way: the fraction of areal exposure that is available as a function of magnitude. This shows that there is almost no loss of areal exposure for long timescale transient before magnitude $R=19.5$, but for the short timescale transients, the areal exposure already starts to decrease at $R\approx18$. This shows that the areal exposure on short timescales is more sensitive to low limiting observations than the longer times scales.  
In the rest of the paper, we will use the areal exposure assuming that all images can be used (the black line in the mid-right panel, 1 in the bottom panel) to calculate the observed rate of transients.

\subsection{Transients}\label{sec:transients}

\definecolor{Gray}{gray}{0.9}

\begin{table*}
\centering
\caption{Overview of all transients found, sorted by discovery date. Shaded rows indicate extragalactic transients. The first detection column lists the time when the transient was first detected on an image. The discovery column shows the time when the transient was identified by the human scanners. The counterpart column lists the magnitude of the quiescent counterpart with $R$ indicating the PTF $R$ magnitude, and $r$ the SDSS or Pan-STARRS $r$ magnitude (the faintest of the two) for point sources. That the counterparts are point-sources has been determined from both the SDSS and Pan-STARRS catalogues. The spectrum column lists the sources of the spectra; 
`ACAM': observations with ACAM at the WHT (see Sec. \ref{sec:surveydesign}),
`R.C. Spec': R.C. Spectrograph at the 4 meter Mayall telescope at KPNO \citep{2010ATel.3027....1A},
`Kast': Kast spectrograph at the 3 meter Shane telescope at Lick Observatory,
`LRIS': LRIS at Keck-1 \citep{1995PASP..107..375O}.
`DBSP': DBSP at the Hale telescope \citep[P200,][]{1982PASP...94..586O}
The superscript above the name refers to one of the following papers:
$a$: \citet{2014MNRAS.438.1391P},
$b$: \citet{2010ATel.3027....1A},
$c$: \citet{2010CBET.2601....1D},
$d$: \citet{2010ATel.3081....1D},
$e$: \citet{2014ApJS..215...14D}, 
$f$: \citet{2014AJ....148...63S},
$g$: \citet{2012MNRAS.426.2359M},
$h$: \citet{2003A&A...412...45P},
$i$: \citet{2014MNRAS.441.1186D},
$j$: \citet{2009PASJ...61S.395K},
$k$: \citet{2015ApJ...807..169A},
$l$: \citet{2017AJ....153..144O},
$m$: \citet{2006A&A...449..425M}.
}
\label{tab:transientoverview}
\begin{tabular}{llllllll}
Name                      & Ra        & Dec       & First detection & Discovery   & Type           & counterpart & spectrum              \\
PTF ...                   & (\degree) & (\degree) & (MJD-55501)     & (MJD-55501) &                &             &                       \\
\hline
10vey                     & 21.734133 & 24.188559 & -46.5           & -46.5       & CV             & $R$=19.37      &                       \\
10aaqh                    & 58.242271 & 21.424262 & -3.76           & -1.78       & M-flare        & $R$=16.50     & ACAM                  \\
\rowcolor{Gray}
10zbk$^a$                 & 37.393757 & 22.333575 & -1.76           & -1.72       & SN/Ia          & galaxy      & R.C. Spec, Kast             \\
\rowcolor{Gray}
10zcd$^b$                 & 17.137369 & 16.502957 & -1.76           & 0.13        & SN/Ia          & galaxy      & ACAM, R.C. Spec           \\
\rowcolor{Gray}
10zej$^{c,d}$             & 31.760443 & 20.853715 & 0.1             & 0.19        & SN/Ia-``91bg''?          & galaxy      & ACAM                  \\
\rowcolor{Gray}
10zeb$^{e}$                     & 26.82027  & 15.82885  & -1.82           & 0.22        & QSO/BL Lac     & $r$=19.20     &                       \\
\rowcolor{Gray}
10zhi$^b$                 & 23.101224 & 21.455835 & 0.19            & 0.27        & SN/Ia          & galaxy      & R.C. Spec                \\
10zfe                     & 30.067678 & 23.926279 & 0.34            & 0.34        & M-flare        & $R$=18.15     &                       \\
\rowcolor{Gray}
10zdq                     & 27.478711 & 21.033048 & 0.19            & 0.36        & SN/Ia          & galaxy      & ACAM                  \\
10zdi$^f$                 & 31.639451 & 20.952067 & 0.1             & 0.44        & CV             & $R$=18.44     & ACAM                  \\
\rowcolor{Gray}
10zdk$^g$                 & 33.530525 & 23.630602 & 0.12            & 0.46        & SN/Ia          & galaxy      & ACAM, R.C. Spec, Kast, LRIS \\
\rowcolor{Gray}
10zqz                     & 25.085293 & 19.205094 & 0.60  & 0.86 & SN & galaxy &  \\
10aacy                    & 27.623838 & 16.107371 & 1.17            & 1.17        & M-flare        & $R$=16.37     & ACAM                  \\
\rowcolor{Gray}
10aadb$^h$                & 35.64689  & 25.137566 & -5.6            & 1.20         & AGN            & galaxy      &  ACAM, DBSP                    \\
10zig$^i$                  & 22.159879 & 18.760014 & 1.31            & 1.31        & CV             & $R$=18.22     &                       \\
10zix$^j$                  & 32.792509 & 17.273396 & 1.16            & 1.33        & CV             & $R$=19.61     & ACAM                  \\
\rowcolor{Gray}
10zxs                     & 37.87864  & 20.461258 & 0.20  & 1.89 & SN & galaxy & \\
10aaop                    & 33.024992 & 25.166939 & 1.1             & 2.18        & M-flare        & $R$=19.51     &                       \\
\rowcolor{Gray}
10aaes$^k$                & 31.791567 & 16.211025 & -19.81          & 2.20         & SN/IIn        & galaxy      & ACAM                  \\
10aaom                    & 54.314202 & 15.914879 & 2.24            & 2.25        & M-flare        & $R$=18.58     &                       \\
\rowcolor{Gray}
10aaho                    & 32.755376 & 15.785523 & 0.14            & 2.27        & SN/IIP         & galaxy      & ACAM, LRIS        \\
\rowcolor{Gray}
10aaey                    & 39.654895 & 22.056594 & 0.29            & 2.31        & SN/Ia          & galaxy      & LRIS                  \\
10aafc$^l$                & 51.512414 & 25.425869 & 2.33            & 2.40         & CV            & $r$=20.17     & ACAM                  \\
10aagv                    & 57.774944 & 21.256824 & 3.34            & 3.34        & M-flare        & $R$=18.34     &                       \\
\rowcolor{Gray}
10aajr$^m$                & 33.219913 & 22.748068 & 0.26            & 4.10         & QSO/BL lac    & $r$=17.83     &                       \\
\rowcolor{Gray}
10aaiw$^a$                & 17.338134 & 15.735881 & -2.71           & 4.13        & SN/Ia-``99T''  & galaxy      & ACAM, KAST             \\
10aakm                    & 51.319163 & 22.735708 & 0.14            & 4.51        & M-flare        & $R$=15.89     &  ACAM                    \\
10aaqc$^i$                & 38.260974 & 18.806173 & 6.28            & 6.28        & CV             & $r$=22.03     & ACAM                  \\
10aarq                    & 28.098845 & 25.613108 & 6.34            & 6.34        & M-flare        & $R$=20.68     &                       \\
10aaqt$^i$                & 23.952331 & 24.400854 & 6.24            & 6.40        & CV             & $r$=23.02     &                       \\
10aaqj                    & 50.600737 & 25.309258 & 5.33            & 6.42        & CV             & $R$=20.16     & ACAM                  \\
10aaqb$^i$                & 59.774695 & 17.842959 & 6.38            & 6.46        & CV             & $R$=18.00     & ACAM                  \\
10aaqu$^i$                & 53.982354 & 19.188484 & 6.36            & 6.52        & CV             & $R$=20.34     &                       \\
1401fi                   & 22.82068  & 26.1559  & 1.29            & >8           & M-flare        & $r$=21.37    &                         \\
\hline
\end{tabular}
\end{table*}

All transients that were identified as real are listed in Table \ref{tab:transientoverview} and shown in Fig.~\ref{fig:FOV}. We found a total of 12 supernovae, 10 cataclysmic variables, 9 flaring M-stars, and 3 flaring AGN. We will discuss each class separately in the following sections.

\subsubsection{Supernovae}
We found a total of 12 supernovae in the Sky2Night area. They are listed in Table \ref{tab:transientoverview} and the light curves and spectra are shown in Fig.~\ref{fig:SNdata}. For most of the supernovae, we have at least 8 nights of photometry. For 10 supernovae, an ACAM spectrum is available. For some supernovae, additional spectra are available that were obtained as part of other programs in the PTF collaboration. For the 2 faint supernovae that were not found in the real-time search (PTF10zqz and PTF10zxs) no spectra are available.

To determine the type and sub-type of the supernovae, we use \textsc{SNID} \citep{2007ApJ...666.1024B} to cross-correlate the spectra with supernova template spectra. If possible, we determine the redshift from the host galaxy or use narrow emission lines in the supernova spectrum. If this is not possible, we use the average redshift from the \textsc{SNID} cross-correlation. For the supernovae without a spectrum, we use the SDSS photometric redshift. To determine the age of the supernovae, we fit a supernova light curve template to the PTF difference imaging photometry using the python package \textsc{sncosmo} \citep{Barbary:11938}. For the Ia supernovae we use the template light curves from \citet{2007ApJ...663.1187H} and for the core-collapse supernovae we use the templates from \citet{1999ApJ...521...30G}\footnote{`Nugent' supernovae templates available at \url{https://c3.lbl.gov/nugent/nugent_templates.html}}.

The results are shown in Table \ref{tab:SNstats}. The majority of the supernovae (8) are of type Ia. Since we obtained both a spectrum and an 8-day light curve, there is little uncertainty in the classification. The subtypes are more difficult to determine, but all except two, appear to be normal type Ia supernovae \citep[for an overview of Ia subtypes, see for example][]{2017arXiv170300528T}. PTF10aaiw, for which we have two spectra, is a ``91T''-like supernova \citep{1992ApJ...384L..15F} according to the cross-correlation with template spectra. The spectra of PTF10aaiw show a shallow Si\,\textsc{ii} 6355\,\AA\ absorption line and deep Fe\,\textsc{iii} absorption features, which are the main features discriminating ``91T''-type supernovae from normal SNIa supernovae. In addition, the absolute peak magnitude as determined from the light curve fit, $M_B\approx -19.5$, is consistent with being a ``91T''-type supernova. PTF10zej could be a ``91bg''-type supernova \citep{1992AJ....104.1543F}; the obtained spectra match almost equally well with ``91bg''-templates and normal Ia-template spectra. The estimated absolute magnitude is only $M_B\approx-18.0$, which puts it at the boundary between normal Ia supernova and ``91bg''-type supernovae. With the available data, we cannot make a certain sub-classification of the subtype of PTF10zej.

PTF10aaho and PTF10aaes, are core-collapse supernovae. PTF10aaho is a supernova that exploded a few days before the start of the program in a faint, unresolved galaxy ($\mathrm{SDSS}_r=21.36$\,mag). The light curve shows a rapid rise during the Sky2Night project, and PTF kept observing the field containing this supernova for a long time. The light curve indicates that this is a normal type IIP supernova \citep[e.g.][]{1997ARA&A..35..309F}.

PTF10aaes is likely a core-collapse supernova that occurred off-centre ($2.4\arcsec$ distance) in an elliptical galaxy. The spectrum best matches to that of type-II SN templates of 80 days or older. The lack of any significant trend in the 8-day light curve and the faint absolute magnitude ($M_B=-15.8\pm0.5$) also indicate that this is most likely an old supernova. However, if PTF10aaes is indeed an old supernova, we should have detected it in the reference images (taken 15 days before the start of the Sky2Night project) and should not have shown up in the difference images. A visual inspection shows a detection in only one out of the 5 individual images used to make the reference image. One possibility is that the supernova re-brightened slightly since the reference images were taken, and therefore does appear in the difference images.

We have been unable to identify two supernovae, PTF10zqz and PTF10zxs. They are transients that appeared close to a galaxy (but not in the nucleus of the galaxy). No spectrum is available for PTF10zqz and PTF10zxs, and the light curve does not show any significant evolution over the 8 days of data. With this information, it is not possible to classify these two supernovae.

We also found one false positive supernova, PTF10zfi, which turned out to be a processing artefact. It appeared close to a galaxy, which is why it was initially confused for a supernova. Two ACAM spectra were obtained of the host galaxy but did not show any sign of a supernova. Re-analysis of the imaging data with forced photometry does not show the transient any more. 


\subsubsection{Outbursting Cataclysmic Variables}
We found a total of 10 outbursting CVs in the Sky2Night survey area, see Table \ref{tab:CV} and Fig.~\ref{fig:CVdata}. Out of these systems, five were found by the human scanners during the Sky2Night project and for these five we obtained an ACAM spectrum. We confirmed that these objects are CVs by their spectra which show Balmer emission lines at a redshift of zero. For the remaining systems, we confirmed the CV nature by inspecting their CRTS light curves, which show many eruptions over the 10+ year baseline. 

Dwarf nova outbursts are the most common type of large amplitude optical variation in CVs, and the majority of CVs we found feature dwarf nova outbursts. The amplitude of the outbursts are typically $R=1-5$\,mag and last approximately 4 to 6 days. Transients PTF10vey, PTF10zdi, PTF10zix, PTF10aafc, and PTF10aaqu are typical examples of dwarf novae outbursts. PTF10aaqb shows an outburst amplitude of only 1.2 magnitudes. However, CRTS archival data show many dwarf nova outbursts with an amplitude of typically 2 magnitudes. We, therefore, conclude that PTF10aaqb is also a dwarf nova outburst.

Transients PTF10aaqc and PTF10aaqt have no counterpart in the PTF images. However, deeper SDSS images and Pan-STARRS images both show a faint, unresolved object. Both transients appeared at the end of the project so the light curve only spans a few days 
and no spectra are available. Both transients are repeating; both in PTF data obtained years later and in CRTS data multiple outbursts of a few days duration can be seen for both objects. We therefore also classify PTF10aaqc and PTF10aaqt as dwarf novae.

Transient PTF10zig was in outburst long enough for it to be detected both in the reference image and during the survey. The PTF light curve shows rapid variability ($\unsim$ hours) of $\unsim$1.5 magnitudes. The few observations were taken before the start of the Sky2Night project hint that this system was already in outburst for 20 days, and possibly 80 days. The CRTS light curve, taken around the same time as the Sky2Night survey, also shows rapid variability of $\unsim$1.5 magnitudes. Observations taken by CRTS years earlier and later only detected the source at $\unsim$21\,mag. The SDSS image also shows a faint source with a colour of $g-r\approx 0.3$ with $r=21.55$. In addition, an SDSS spectrum is available which shows many Balmer emission lines and also the He\,\textsc{I} emission line at $5875\,$\AA\ which confirms the CV nature of the object. 
However, the light curve does not resemble that of a typical CV with dwarf nova outbursts. The PTF observations could be taken while it was in a super-outburst; long outbursts that can last months and can feature strong variability \citep[e.g.][]{2013PASJ...65...50O}. If indeed the Sky2Night light curve is part of a long duration superoutburst, PTF10zig can be classified as an SU UMa or WZ Sge subtype of dwarf nova CVs.

PTF10aaqj showed a slow brightening of about 1 magnitude during the Sky2Night project and was therefore saved as a candidate. The ACAM spectrum feature Balmer emission lines ($z=0$) and confirms that this is a CV. The CRTS light curve shows non-periodic optical variability but with no clear outbursts. These characteristics match those of AM Her type CVs \citep{2003cvs..book.....W}.

\subsubsection{Flaring M-stars}
A total of 9 flaring stars were identified in the Sky2Night data, see Fig.~\ref{fig:flaredata}. All except PTFS1401fi were identified as candidate transients, and a spectrum was obtained for three of the objects. The quiescent counterparts of the flaring objects were also detected in PTF reference images, ranging from $R=16$\,mag to $R\approx 21$\,mag. We use Pan-STARRS colours to determine the spectral type of the M-dwarfs, 
following the classification of \citet{2018ApJS..234....1B}, see Table \ref{tab:dMe} and Fig. \ref{fig:PScolours}. This shows that the majority of the flaring objects are of spectral type M4--M5. Two of the flaring stars were significantly redder and have later spectral types of M6 and M7.

We fit a simple outburst model (instant rise, exponential decay in flux) to the light curves to determine the outburst properties, such as flare magnitude and decay time scale, see Table \ref{tab:dMe}. Here we have assumed that the highest detected magnitude corresponds to the observed peak magnitude, the most conservative approach. We calculated the energy emitted per flare in the $R$-band by first measuring the equivalent duration of the flare \citep{1972Ap&SS..19...75G}, and then calculate the absolute energy in the flare by using the absolute magnitudes of M-stars from \citet{2013ApJS..208....9P}. 

The flare timescales are typically within 0.5 and 2 hours, with one longer flare with a timescale of almost 5 hours. The observed flare magnitudes are typically between 0.6 and 1.5 magnitudes, but three flares are significantly stronger, with the strongest flare of 3.5 magnitudes.

\subsubsection{AGN activity}
Three promising transient candidates were followed up with the WHT but were identified as an AGN. Their light curves and spectra are shown in Fig.~\ref{fig:otherdata}.

PTF10aadb is a transient at the core of a face-on SAa spiral galaxy \citep[$z\approx0.062$,][]{2012ApJS..199...26H}. The light curve during the Sky2Night project has an average of $R\approx 19.5$\,mag and does not show a significant trend. The initial spectrum showed what seemed to be a weak H$\alpha$ P-Cygni profile, and the object was initially identified as a type IIn supernova. However, there is evidence that the core of the galaxy is an AGN. First, a radio source has been detected in the NVSS survey \citep{1998AJ....115.1693C}. Second, observations by PTF three years after the Sky2Night project also show another brightening of the core of this galaxy, now up to $R\approx 18.8$\,mag. This makes it more plausible that the transient seen during Sky2Night is due to AGN activity. A spectrum obtained 25 days after the first spectrum shows a huge increase in $H\alpha$ emission and that the O-III lines (4959 and 5007\AA) disappeared. In addition, the AGN became significantly brighter at the long wavelengths (>7000\AA). The strong increase of $H\alpha$ emission is a typical characteristic of changing look AGN \citep[e.g.][]{2017ApJ...835..144G}. 

PTF10zeb and PTF10aajr both appear as unresolved sources which rapidly became brighter. For PTF10aajr a spectrum was obtained with ACAM. The spectrum shows a blue continuum without any prominent features. Both sources are known radio and X-ray sources and classified as BL Lac-type objects \citep{2006A&A...449..425M,2014ApJS..215...14D}, which agrees with our observations.

\subsection{Observed transient rates}\label{sec:rates}


\begin{table}
\caption{The observed rate of transients for the Sky2Night project. The uncertainties indicate the 95 per cent confidence interval. Upper-limits for fast optical transients are 95 per cent confidence upper-limits. These limits assume a detection efficiency of $\epsilon=1$ (see equation \ref{eq:rates}) and assume that transients could have been detected in all of the images.}
\label{tab:rates}
\setlength\extrarowheight{6pt}
\begin{tabular}{lll}
Type           & $N$ & $\mathcal{R}$ ($10^{-4}$ \psdd) \\
\hline
Supernova - Ia & 8 & $\phantom{00}10.0^{+8.2}_{-5.3}$
 \\ 
Supernova - CC & 1 & $\phantom{000}2.0^{+4.4}_{-1.6}$ \\  
CV - DN  & 5 & $\phantom{00}12.8^{+13.6}_{-7.9}$ \\  
CV - DN - $R>20.5$\,mag & 2 & $\phantom{000}6.0^{+10.3}_{-4.6}$ \\ 
M-flares       & 9 & $\phantom{0}118\phantom{.0}^{+94}_{-58}$ \\
M-flares - $R>20.5$\,mag     & 2 & $\phantom{00}35\phantom{.0}^{+64}_{-26}$ \\
BL Lac flares  & 2 & $\phantom{000}6.7^{+11.6}_{-5.0}$ \\
FOTs (4\,h)    & 0 & $<37$ \\
FOTs (1\,d)    & 0 & $<\phantom{0}9.3 $ \\
\hline
\end{tabular}
\end{table}

The observed rate of transients is calculated as follows:
\begin{equation} \label{eq:rates}
    \mathcal{R} = \dfrac{N}{\epsilon^2 E_A (\tau)}\ (\mathrm{deg}^{-2}\,\mathrm{d}^{-1})
\end{equation}
with $N$ the number of transients, $\epsilon$ the detection efficiency per image, and $E_A$ the effective exposure (see Sec.~\ref{sec:observations}). Since $N$ is a small number, we use Poisson statistics to calculate the uncertainty \citep[e.g.][]{Gehrels:1986}. 

We used a simple estimate for the detection efficiency: all transients brighter than the detection limit are recovered ($\epsilon=1$) and those fainter than the detection limit are not ($\epsilon=0$). The detection efficiency ($\epsilon$) occurs in the equation squared because we required a transient to be detected in two images. The efficiency is difficult to estimate and is a function of the magnitude, the background (e.g. a galaxy), and the subjective nature of human scanners. We tried to be as complete as possible by saving candidate transients when in doubt. However, the efficiency will always gradually decrease as the brightness approaches the detection limit. \citet{2017ApJS..230....4F} performed a detailed test of the recovery rate as a function of limiting magnitude, brightness of the transient, seeing, angular distance to the nearest galaxy and other parameters. Such a level of detail is not needed in this work, since the Poisson uncertainty dominates the rates and is of the order of 20 per cent or more. We note that \citet{2017ApJS..230....4F} found a maximum recovery efficiency of 97 per cent. In the calculation of the rates, we will assume $\epsilon=1$ for transient brighter than the magnitude limit. This should be kept in mind when interpreting the results: if the real efficiency is lower than 1, the real observed transient rates will be slightly higher than reported in this work.

The effective exposure ($E_A$) depends on the visibility time of the transient, as can be seen in Fig. Fig.~\ref{fig:survey_characteristics}. We, therefore, need to estimate how long a transient would have been visible during the project. We then use the areal exposure assuming that the transients could have been detected in all images (the black line in the bottom-left panel in Fig.~\ref{fig:survey_characteristics}). For supernovae, we assume that they are all visible for longer than 15 days. This maximum results from the requirement of a non-detection in the reference images obtained 15 days before the start of the project. For the dwarf novae and BL Lac flares we estimate the visibility time by eye from the light curves, ranging between 3--5\,d. For the M-dwarf flares, we have used the fitted curve to estimate the visibility time, which are typically detectable as transient for 3-6\,h. We assume an uncertainty on our estimates of the visibility time ($\tau$) 10\% (log-normal distributed).

We calculate the observed rate and uncertainty for each type of transient by numerically combining the Poisson distribution for $N$, with the distribution we calculated for $E_A$. The final values are shown in Table \ref{tab:rates} with the uncertainties indicating the 95~per cent confidence interval. In addition, we calculate an upper limit for transients visible for 4\,h and 1\,d. We use the 95~percentile upper limit, which corresponds approximately to 3~detections \citep{Gehrels:1986}.

\section{Discussion}\label{sec:discussion}
\subsection{Expected number of transients}
\subsubsection{Supernovae}
The expected number of supernovae Ia in the survey are easy to determine since their light curves are very uniform and the volumetric rate for Ia supernovae is well-known \citep[e.g. ][]{2014ApJ...783...28G}. In addition, they can be assumed to be uniformly distributed across the sky. We use \textsc{sncosmo} to simulate a large number of supernova Ia light curves of uniformly distributed supernovae (in co-moving volume). We use SALT2 supernova light curve templates \citep{2007A&A...466...11G} with parameters and host galaxy extinction parameters according to \citet{2017ApJ...842...93M}. We also take into account Galactic extinction \citep{1998ApJ...500..525S}. We simulate our survey by checking if the supernovae are in the Sky2night field and if it is above the detection limit during the survey and not detected in the reference image. For a limiting magnitude of $R$=20.21\,mag (the magnitude at which the fractional areal exposure is 90 per cent, see the bottom panel of Fig. \ref{fig:survey_characteristics}) and volumetric rate of $\mathcal{R_\mathrm{volume}}=2.5\times 10^{-5} \mathrm{yr^{-1} Mpc^{-3}}$ \citep{2014ApJ...783...28G}, we would expect to find 13.4 supernovae during our project, marginally consistent with 8 confirmed detections taking into account Poisson uncertainty. Or, the other way around, we find a volumetric rate of $\mathcal{R_\mathrm{volume}}=1.50_{-0.73}^{1.46} \times 10^{-5} \mathrm{yr^{-1} Mpc^{-3}}$. This suggests that we have missed some supernovae in our search or the effective magnitude limit was about $R\approx 19.8$\,mag. \citet{2018PASP..130c4202A}, who also used PTF data, also found that the efficiency started to drop of about 0.5 magnitudes above the detection limit. This could be explained due to the fact that Ia supernovae occur in galaxies, which makes their detection more difficult. 

We compare the relative number of different types of supernovae to the fraction of supernovae found by the Lick Supernova search \citep{2011MNRAS.412.1441L}. They find that 74 per cent of their supernovae are of type Ia, with 17 per cent of type II and 9 per cent type Ibc. Our results are consistent with this result: with 8 out of 9 identified supernovae are of type Ia (we do not count PTF10aaes, as it is an old supernova). The distribution of Ia subtypes is also in agreement with the ratios found in the Lick Supernova search. With 8 detected Ia supernovae, the expectation value for ``99T'' and ``91bg'' types is both one, which is what we have found.

\subsubsection{Dwarf novae}
\begin{figure}
\centering
  \includegraphics{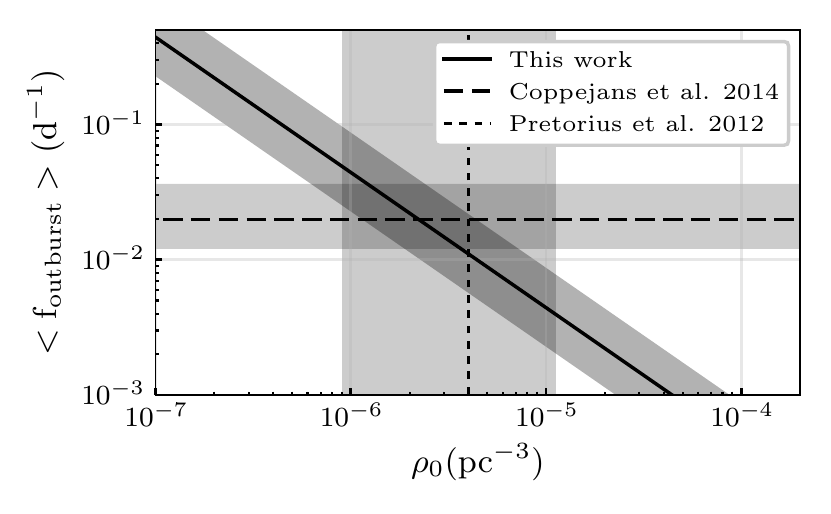}
 \caption{The CV space density versus the average dwarf nova outburst frequency. The dwarf nova volumetric rate we derived in this paper is shown as the diagonal line in the figure. The vertical line shows the space density as measured by \citet{2012MNRAS.419.1442P} and the horizontal line shows the average outburst frequency calculated from the dwarf nova sample by \citet{2014MNRAS.437..510C}. The shaded regions indicate a 95 per cent confidence interval.}
  \label{fig:DNflux}
\end{figure}

We simulate the number of dwarf novae outbursts which we expect given the CV space density and outburst frequency. We use a simple Galaxy model with a thin and thick disk. We assume disks with an exponentially decreasing density profile with scale heights of 200\,pc and 1000\,pc and a Galaxy scale radius of 3000\,pc \citep{2004MNRAS.349..181N}. We assume a density ratio between the thin and thick disk population of cataclysmic variables of 1 to 56 (Groot et al., in prep.). We randomly populate the model Galaxy according to the space density distribution and keep only the objects that are in the field-of-view of the Sky2Night survey. We then estimate if each object would have been detected in the Sky2Night project if a dwarf nova outburst occurs. For this, we use the relation $M_V=5.92-0.383P$, with P as the orbital period \citep{1987MNRAS.227...23W}, and assume that \mbox{$V-R$ = 0} at peak. We randomly draw periods from a sample of periods as found in CRTS \citep[see][]{2014MNRAS.437..510C}. With this simulation and 8 detected dwarf novae, we derive a local dwarf nova volumetric rate of $4.6_{-2.4}^{4.7} \times 10^{-8}\,\mathrm{d^{-1} pc^{-3}}$. As this rate is the combination of a space density and an outburst frequency, we can compare our result to measurements of either of these, which is illustrated in Fig. \ref{fig:DNflux}. This shows that our finding is consistent with the combination of the measured space density of $2.3^{9.0}_{-1.4} \times 10^{-6}\,\mathrm{pc^{-3}}$ \citep[approximated 95 per cent interval,][]{2012MNRAS.419.1442P} and, at the same time, an average outburst frequency of dwarf novae of $20^{+15}_{-8} \times 10^{-3}\,\mathrm{d^{-1}}$ \citep[calculated from the sample of][]{2014MNRAS.437..510C}.

\subsubsection{Stellar flares}
\begin{figure}
\centering
  \includegraphics{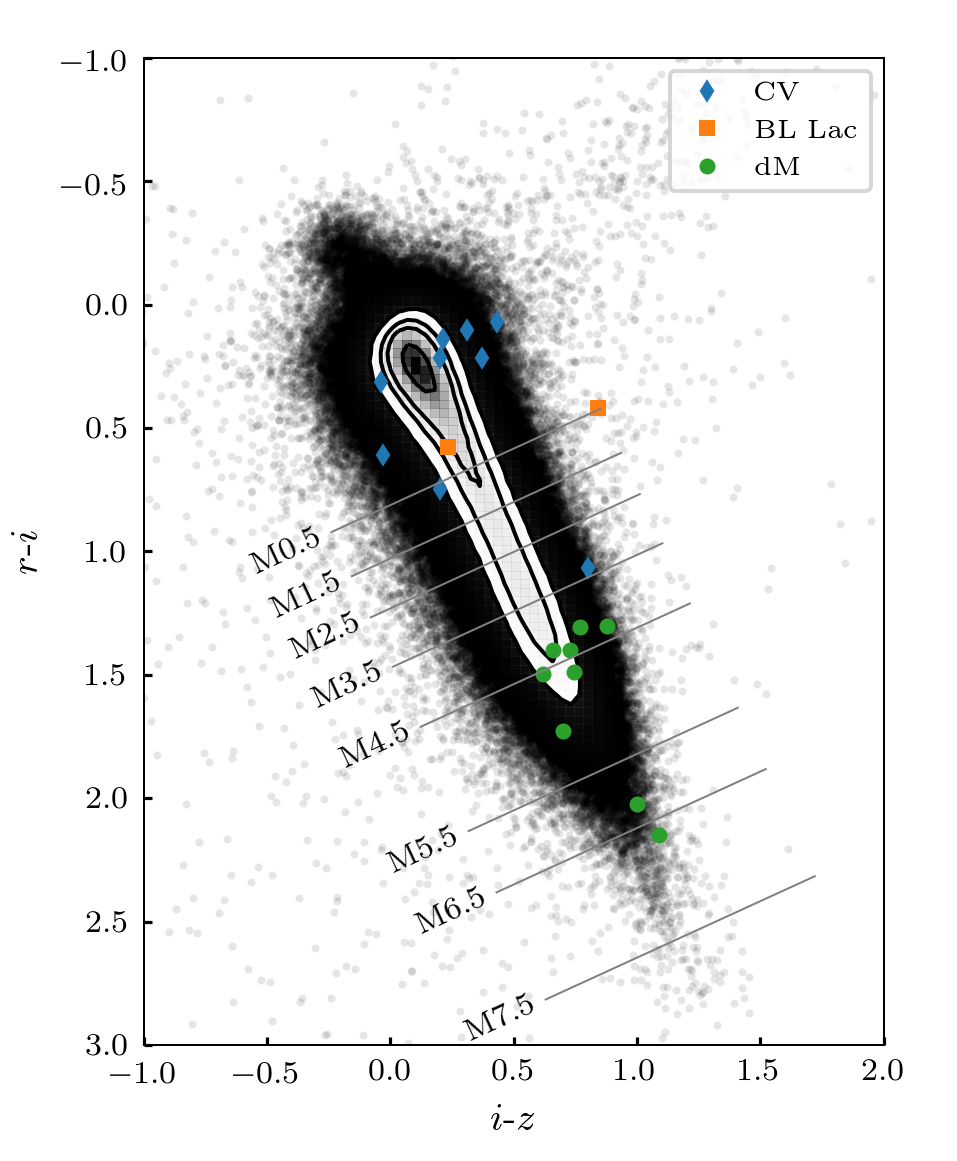}
 \caption{The Pan-STARRS colours of unresolved counterparts of transients. The black dots and contours show all point-sources in the Sky2Night area. 
}
  \label{fig:PScolours}
\end{figure}

\begin{figure}
\centering
  \includegraphics{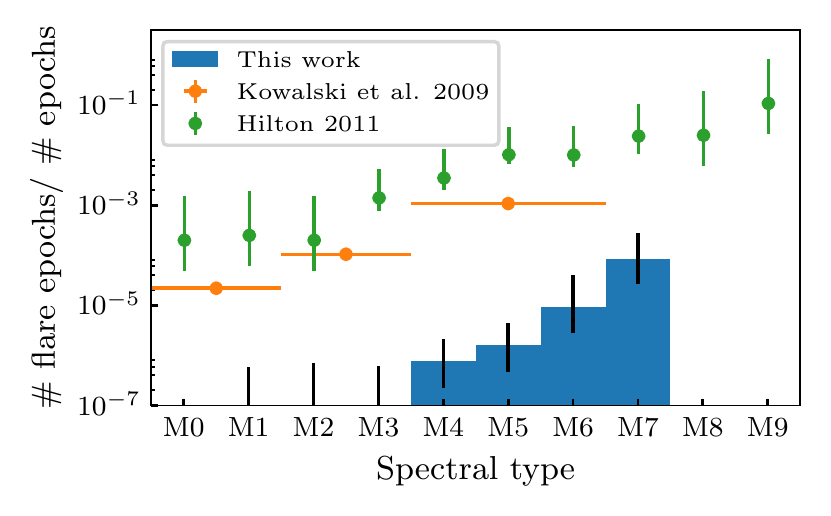}
 \caption{The fraction of time M-stars are in outburst for different spectral types. This is calculated by counting the number of flares per spectral type and dividing by the total number of M-stars of that spectral type in the S2N field, see Fig. \ref{fig:PScolours}. The error bars indicate the Poisson uncertainties only. 
 }
  \label{fig:flareactivity}
\end{figure}

In order to make a more direct comparison with flare rates from other surveys, we calculate the average flare duty cycle per M-dwarf spectral type. First, we use $ri$ and $iz$ colours from Pan-STARRS \citep{2016arXiv161205243F} to determine the number of stars per spectral type in the Sky2Night area, see Fig. \ref{fig:PScolours}. To calculate the average flare duty cycle we divide the number of flare epochs by the total number of observations per spectral type, see Fig. \ref{fig:flareactivity}. 

This shows that, on average, the late type M-dwarfs are more active, which confirms earlier findings by \citet{2009AJ....138..633K} and \citet{2010AJ....140.1402H} (both with SDSS data), also plotted in Fig. \ref{fig:flareactivity}. All findings show a similar trend, but the absolute numbers are three orders of magnitude off. This is the result of different flare selection criteria: \citet{2009AJ....138..633K} selected flares with $\Delta u>0.7$ in Stripe 82 data, and \citet{2010AJ....140.1402H} use Balmer emission lines in SDSS spectra to identify flares, but these lines can be a sign of persistent chromospheric activity as well. The main difference is that the contrast in the $u$-band and of emission lines of flares is much higher than in the $R$-band. Models by \citet{2012ApJ...748...58D} can be used to convert $\Delta R$ to $\Delta u$ (assuming $r=R$); a flare of $\Delta R = 0.6$ on an M4 star corresponds to a $\Delta u$ of 4 magnitudes. This makes all flares which we found brighter than at least 99 per cent of the flares found in \citet{2009AJ....138..633K}. This explains the large difference in observed rates: \citet{2009AJ....138..633K} reported an observed rate of 48 flares\,\psdd, a factor 2600 higher than our results (see Table \ref{tab:rates}). This is also consistent with the difference between the duration and flare energy compared to the relation plotted in Fig. 5.8 of \citet{2011PhDT.......144H}. The flares found in the Sky2night project are all at the high-energy, long-decay time end of the distributions.

We compare the rate of flares with the 38 M-flares found in the entire iPTF survey by \citet{Ho2018}. Using their estimate for $E_A=22146$\sdd, the rate of such flares is $\mathcal{R}=17^{+6}_{-5} \times 10^{-4}$\,\psdd. \citet{Ho2018} rejected any transients with a stellar counterpart in the PTF reference images, so we compare it to the rate of flares with a counterpart fainter than the detection limit: $35^{+64}_{-26}\times 10^{-4}$\,\psdd (Table \ref{tab:rates}). The rate from Sky2Night is slightly higher but consistent with the flare rate by \citet{Ho2018}.

\subsection{Upper limit for fast optical transients}\label{sec:fots_limits}
\begin{figure*}
  \centering
  \includegraphics{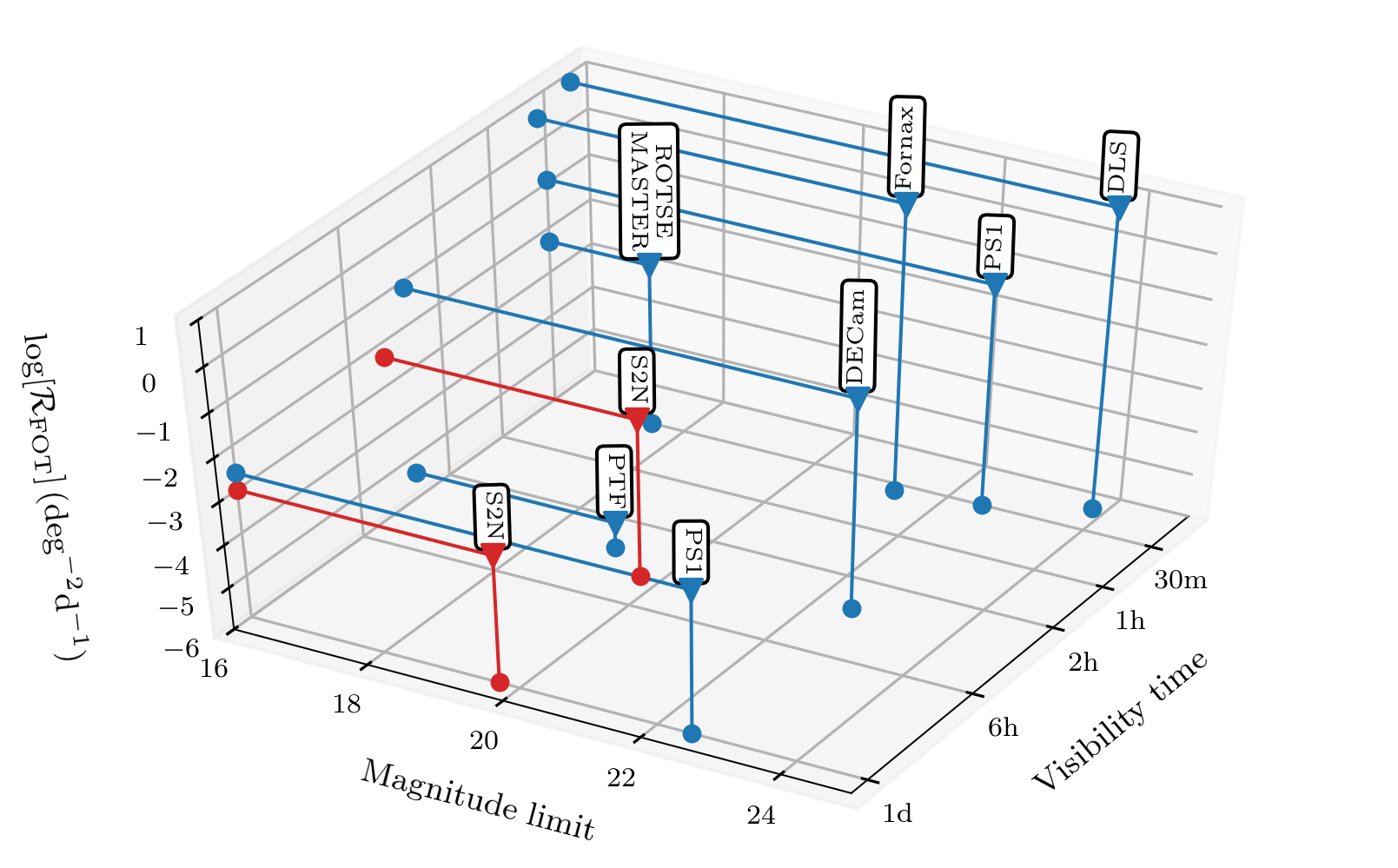}
 \caption{Upper-limit to the rate of fast optical transient versus time-scale and limiting magnitude \citep[adapted from][]{2013ApJ...779...18B}. The limits set by this work are shown in red and are labelled `S2N'. Upper-limits by other surveys are indicated in blue; Deep Lensing Survey \citep[DLS,][]{2004ApJ...611..418B},
 ROTSE/MASTER \citep{2005ApJ...631.1032R,2007ARep...51.1004L}, 
 Fornax Survey \citep[Fornax,][]{2008ApJ...682.1205R}, 
 Pan-STARRS \citet[`PS1',][]{2013ApJ...779...18B}, 
 survey by the DECam \citep[DECam,][]{2017arXiv171002144C},
 and a search of all PTF data \citep[PTF,][]{Ho2018}.
 }
  \label{fig:FOTSrate}
\end{figure*}

Since no unclassified fast optical transients were found in our search, we have calculated upper limits for the rate of \fots\ visible for 4 hours and 1 day (see Sect. \ref{sec:rates} and Tables \ref{tab:rates}).
We compare our upper limits to upper limits determined by other searches for \fots, see Fig. \ref{fig:FOTSrate}. Our result is most similar to the upper limit set by \citet{2017arXiv171002144C}: 0.07\,\psdd\ down to 22.5 in $i$-band at a time scale of 3 hours. The Sky2Night upper limit is a factor of $\approx 15$ lower, but at magnitude $19.7$, 2.8 magnitudes lower. The Sky2night upper limit for 1\,d transients is a factor 2.5 times deeper than the limit set by \citet{2013ApJ...779...18B} using $g$- and $r$-band data from Pan-STARRS. However, the PanSTARRS limiting magnitude is again 2.8 magnitudes deeper than the Sky2night search, making the Pan-STARRS upper limit slightly more constraining.

A lower limit to the rate of \fots\ is set by GRB afterglows. During the entire duration of PTF, one GRB afterglow was found as a fast optical transient: PTF14yb \citep{2015ApJ...803L..24C}. The transient was bright enough to be detected by PTF for a total of five hours. \citet{Ho2018} did an archival search of all PTF transients and did not find any new \fots\ besides flaring M-dwarfs. Given this one event, they calculated a rate for extragalactic \fots\ (peak $m \leq 18$ and fade by $\Delta \mathrm{mag} > 2$ in $\Delta t=3$\,hrs) of $\mathcal{R}=4.5_{-4.4}^{+17.8} \times 10^{-5}$\,\psdd\, \citep[see also ][]{2015ApJ...803L..24C}. This indicates that the limit set by the Sky2Night survey is approximately 2 orders of magnitude above the rate of extragalactic fast optical transients.

\begin{figure}
\centering
  \includegraphics{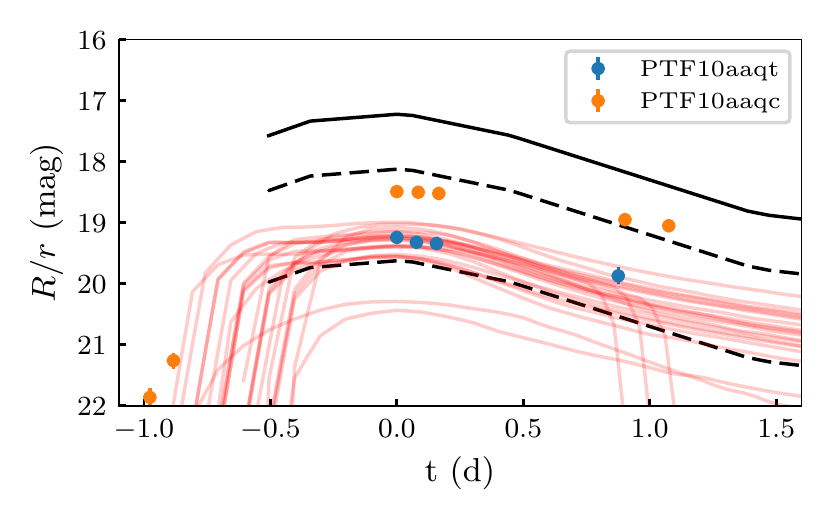}
 \caption{A comparison between the light curves of the DN detected in the Sky2Night project and kilonova light curves. All light curves have been shifted such that their peak occurs at t=0. The black line indicates the best fit light curve of AT2017gfo in $r$-band \citep{2017ApJ...851L..21V}. The dashed and dotted lines show this light curve for distances of 60\,Mpc and 120\,Mpc, the expected range of aLIGO/aVirgo during `O3'. The red lines show kilonovae models by \citet{2017Natur.551...80K} at a distance of 120\,Mpc.}
  \label{fig:KNlc}
\end{figure}

\subsection{False positives in the search for kilonovae}
The aLIGO/aVirgo detectors are scheduled for another observing run, starting in early 2019 (``O3''). The estimated distance horizon to detect BNS mergers is 65-120\,Mpc, and the expected number of BNS detections is 1 to 50 events \citep{2016LRR....19....1A}. Systematic follow-up of all BNS gravitational wave events allows us to study kilonovae in more detail using spectroscopy and also determine the characteristics of the population of kilonovae. In order to do this, optical survey telescopes need to quickly identify the kilonova counterpart in an area of 120-180\sd\  \citep{2016LRR....19....1A}. This will be more challenging than the search for the optical counterpart to GW170817, AT~2017gfo, which was well-localized (40\,\sd), nearby ($\approx 40$\,Mpc) and bright ($r\approx17$\,mag at peak). For nearby kilonovae, a galaxy-targeted survey is more efficient than surveying the entire aLIGO/aVirgo error-box \citep[e.g.][]{2016ApJ...820..136G}. However, because aLIGO/aVirgo will be more sensitive, the majority of the BNS events will be more distant and thus fainter. In those cases, a Galaxy targeted search is less viable as the nearby Galaxy census is less complete at higher distances \citep[e.g.]{2018ApJ...860...22K}.
Instead, an untargeted search will be needed to locate kilonova counterparts. The different search strategy, but also the fainter target, means that the number of false positives can become problematic. False positives delay the identification of the true kilonova counterpart and ruling them out requires valuable follow-up resources. In this section, we use results from the Sky2Night project to assess how problematic false positives are in a monochromatic kilonova search, and determine the best way to recognise false positives.

The Sky2Night survey area of the 407\,\sd, 2--3 times the typical ``O3'' errorbox of aLIGO/aVirgo, contained a total of 1012 transient candidates. Most of these were associated with variable stars or bad subtractions of stars (873 out of 1012). These can be identified by the presence of a star in the reference images (or other surveys). During the execution of Sky2Night, the identification of stellar counterparts was done by human inspection, but an automated procedure is easy to implement and is now standard in many transient identification pipelines \citep[e.g.][]{2017AJ....153...73M}. In addition, better image subtraction techniques such as ZOGY \citep{2016ApJ...830...27Z} and also more advanced `RealBogus' methods \citep[e.g.][]{2017MNRAS.472.3101G} have been developed. Moreover, the increase in available training data for the machine learning based `RealBogus' also improves the identification of real transients.
Improvement of this processing step will reduce, in particular, false positives due to poor subtractions of images, and should especially help in removing any nuclear transients which are the result of a slight misalignment of images. If we reject any candidate with a point-source counterpart (888), which is an improper subtraction (26), which is located at the core of a galaxy (48), or is moving (35), only 15 real transients remain out of the 1012 initial candidates.

The remaining 15 transients either have a nearby galaxy as a counterpart or no counterpart at all in the PTF images. The time evolution of a transient is one of the most discriminating properties of a transient, and Sky2Night light curves probe the evolution on timescales of 2 hours to 8 days. If we compare the light curves of the flare stars with that of a kilonova, we can easily tell them apart as flares evolve much faster than any kilonova. On the other hand, kilonovae evolve significantly in a timespan of 8 days, while supernovae lightcurves generally change only slightly in an 8-day timespan. They are therefore also easy to distinguish with an 8-day light curve. However, the two outbursting CVs with a faint quiescent counterpart (PTF10aaqc and PTF10aaqt) evolve on a similar timescale as kilonovae, as shown in Fig. \ref{fig:KNlc}. The rise time is $<$1\,d and the decay timescale is $\approx$1\, mag\, d$^{-1}$. Although the dwarf novae in our example rise and decay slightly slower, the difference will be almost impossible to detect with just a few epochs of data. With only this short span of data and if no additional information, such as historic light curves that may identify previous dwarf nova outburst, making the distinction between a kilonova and a young dwarf nova will be problematic. 
To distinguish a kilonova and a dwarf nova we need to rely on additional information (if no counterpart is detected). In the case of PTF10aaqc and PTF10aaqt, PTF detected previous outbursts, which confirmed the dwarf nova nature of the transients.

With the complete light curves, we have been able to reject all transients as kilonova candidates. However, the goal of kilonova searches is to identify the kilonova as fast as possible so it can be targeted for follow-up observations. Therefore, we should consider the question: can we reject all transients with only one day of data? In that case, the transient light curves contain only a few epochs, and the time-span is only a few hours. Next to dwarf novae now also supernovae become problematic, depending on the time since the merger for the kilonova searches. Supernovae do not evolve significantly in an 8-hour timespan, and therefore can be confused with a kilonova at peak when it is relatively constant (approximately 10-24 hours after the merger, $\approx\!24$ hours for AT~2017gfo). This ambiguity can be solved if the host redshift is known, and the absolute magnitude of the transient can be calculated. However, this information is not always available at the moment the transient is first detected.


We conclude that the rapid (within 24 hour) unique identification of (faint) kilonovae using only a monochromatic transient light curve is difficult due to false positives. This is especially a problem if no counterpart can be identified and no recent pre-merger images are available. The latter issue can be solved by monitoring the entire (available) sky every night. This ensures that all repeating Galactic dwarf novae will be discovered and also that slowly evolving supernovae can be identified and can be discarded as kilonovae. However, this is resource intensive and not always possible (due to weather) and would still leave infrequent outbursting dwarf novae and young supernovae as potential false positives. The second solution is to obtain additional information by performing a two-band survey search to obtain instantaneous colours (e.g. $g-r$) of all transients. According to simulations kilonovae rapidly become redder within the first few days. This was also seen for AT2017gfo were the $g-r$ colour changed by $\approx 0.8\,\mathrm{mag\,d^{-1}}$. This means that within 8 hours it became redder by $\approx0.3\,\mathrm{mag}$ which should be easily detectable, even at low signal-to-noise ratios. 


\citet{2017arXiv171002144C} have explored the colour solution and performed an empirical study of false positives with DECam (mounted at the 4\,m Blanco telescope). They surveyed an area of 56\sd\ for 5 nights at a cadence of 3\,h with $i-z$ colour information. Out of the 929 transient candidates they found, all but 21 can be rejected as a potential kilonova using static colour information. A further inspection of the luminosity and colour evolution is enough to reject all of them as kilonova candidates.
\citet{2018PASJ...70....1U} empirically tested the number of false positives expected in searches for kilonovae with the Hyper-SuprimeCam installed on the 8.2\,m Subaru telescope. They obtained two sets of $i$ and $z$ images separated by 6 days covering 64\,\sd, in which they discovered a total of 1744 transient candidates. They applied colour and variability cuts aimed at identifying kilonovae. They concluded that all supernovae and AGN can be rejected as kilonovae candidates. However, two transients remained satisfying the kilonova criteria: a flare on a M-dwarf or M-giant and a CV outbursts.

The conclusion is that the automatic rejection of all false positives without any additional follow-up is difficult. We have shown that high-cadence (2\,h) survey alone is not sufficient to identify all transients. A combination of high cadence, multicolour light curves combined with historical information are needed to quickly identify transient found in gravitational wave follow-up. We note that the biggest colour change is between the extremes of the optical regime (faster decay with bluer colour).  Therefore a colour such as ($u-z$) would have the highest diagnostic power when probing deep enough. 

\section{Summary and conclusions}
In this paper, we present a systematic, unbiased survey of intra-night transients. We used PTF to survey 407\,\sd\ at a cadence of 2 hours combined with large-scale, systematic follow up with the WHT telescope. We performed a thorough search for transients, both Galactic and extragalactic. Our search identified 35 transients: 8 type-Ia SN, 2 Core-collapse SN, 3 unknown SN, 10 outbursting CVs, 9 flaring M-stars and 3 AGN flares. For each of these types of transients, we have calculated an observed rate and confirmed these with simulations. We found no extragalactic \fots\ and set a deeper upper limit on their observed rate.

Our main conclusions are that the rate of fast extragalactic transients is low, $\mathcal{R}<37\times 10^{-4}$\,\psdd\ and $\mathcal{R}<9.3\times 10^{-4}$\,\psdd\ for timescales of 4\,h and 1\,d at a limiting magnitude of $R\approx19.7$., and that they are not a source of confusion when searching for kilonovae. In addition, a mono-chromatic survey with a cadence of 2 hours, combined with longer time baseline information and static colour information is sufficient to be able to identify common transients such as flaring star, outbursting CVs and supernovae. Difficulties arise if the transients need to be identified within a single night, with only single-band photometry. Transient surveys that aim to identify kilonovae within the first night should observe with at least two bands, preferably widely separated in wavelength, multiple times per night.

\section*{Acknowledgement}
We thank the referee for thoroughly reading
the manuscript and providing us with useful comments and
suggestions.

JvR acknowledges support from the Netherlands Research School of Astronomy (NOVA) and Foundation for Fundamental Research on Matter (FOM), and also the California Institute of Technology where a large part of this work was conducted.

Based on observations made with the William Herschel Telescope (WHT) operated on the island of La Palma by the Isaac Newton Group in the Spanish Observatorio del Roque de los Muchachos of the Instituto de Astrof\'{i}sica de Canarias. 

The Palomar Transient Factory project is a scientific collaboration between the California Institute of Technology, Columbia University, Las Cumbres Observatory, the Lawrence Berkeley National Laboratory,
the National Energy Research Scientific Computing Center, the
University of Oxford, and the Weizmann Institute of Science.

We thank E.~Hsiao, N.~Suzuki, and J.~Botyanszki for the spectra of PTF10zbk and PTF10zdk observed with the Lick telescope. We thank I.~Arcavi, D.~Xu, and T.~Matheson for observing PTF10zbk and PTF10zdk with the KPNO Mayall telescope. We thank E.~Hsiao, N.~Suzuki, J.~Botyanszki and B.~Cenko for the LRIS spectra of PTF10zdk, PTF10aaho, and PTF10aaey.

This research has made use of the SIMBAD database, operated at CDS, Strasbourg, France.

This research made use of Astropy, a community-developed core Python package for Astronomy \citep{2013A&A...558A..33A} 

Funding for the Sloan Digital Sky Survey IV has been provided by the Alfred P. Sloan Foundation, the U.S. Department of Energy Office of Science, and the Participating Institutions. SDSS-IV acknowledges support and resources from the Center for High-Performance Computing at the University of Utah. The SDSS web site is www.sdss.org. SDSS-IV is managed by the Astrophysical Research Consortium for the Participating Institutions of the SDSS Collaboration including the Brazilian Participation Group, the Carnegie Institution for Science, Carnegie Mellon University, the Chilean Participation Group, the French Participation Group, Harvard-Smithsonian Center for Astrophysics, Instituto de Astrof\'isica de Canarias, The Johns Hopkins University, Kavli Institute for the Physics and Mathematics of the Universe (IPMU) / University of Tokyo, Lawrence Berkeley National Laboratory, Leibniz Institut f\"ur Astrophysik Potsdam (AIP), Max-Planck-Institut f\"ur Astronomie (MPIA Heidelberg), Max-Planck-Institut f\"ur Astrophysik (MPA Garching), Max-Planck-Institut f\"ur Extraterrestrische Physik (MPE), National Astronomical Observatories of China, New Mexico State University, New York University, University of Notre Dame, Observat\'ario Nacional / MCTI, The Ohio State University, Pennsylvania State University, Shanghai Astronomical Observatory, United Kingdom Participation Group, Universidad Nacional Aut\'onoma de M\'exico, University of Arizona, University of Colorado Boulder, University of Oxford, University of Portsmouth, University of Utah, University of Virginia, University of Washington, University of Wisconsin, Vanderbilt University, and Yale University.  

The Pan-STARRS1 Surveys (PS1) have been made possible through contributions of the Institute for Astronomy, the University of Hawaii, the Pan-STARRS Project Office, the Max-Planck Society and its participating institutes, the Max Planck Institute for Astronomy, Heidelberg and the Max Planck Institute for Extraterrestrial Physics, Garching, The Johns Hopkins University, Durham University, the University of Edinburgh, Queen's University Belfast, the Harvard-Smithsonian Center for Astrophysics, the Las Cumbres Observatory Global Telescope Network Incorporated, the National Central University of Taiwan, the Space Telescope Science Institute, the National Aeronautics and Space Administration under Grant No. NNX08AR22G issued through the Planetary Science Division of the NASA Science Mission Directorate, the National Science Foundation under Grant No. AST-1238877, the University of Maryland, and Eotvos Lorand University (ELTE). 

This research made use of matplotlib, a Python library for publication quality graphics \citep{Hunter:2007}






\bibliographystyle{mnras}
\bibliography{bibfiles/S2N_misc,bibfiles/S2N_FOTS,bibfiles/S2N_transient_refs,bibfiles/S2N_KNdiscovery,bibfiles/S2N_KNmodels} 

\begin{thebibliography}{}
\makeatletter
\relax
\def\mn@urlcharsother{\let\do\@makeother \do\$\do\&\do\#\do\^\do\_\do\%\do\~}
\def\mn@doi{\begingroup\mn@urlcharsother \@ifnextchar [ {\mn@doi@}
  {\mn@doi@[]}}
\def\mn@doi@[#1]#2{\def\@tempa{#1}\ifx\@tempa\@empty \href
  {http://dx.doi.org/#2} {doi:#2}\else \href {http://dx.doi.org/#2} {#1}\fi
  \endgroup}
\def\mn@eprint#1#2{\mn@eprint@#1:#2::\@nil}
\def\mn@eprint@arXiv#1{\href {http://arxiv.org/abs/#1} {{\tt arXiv:#1}}}
\def\mn@eprint@dblp#1{\href {http://dblp.uni-trier.de/rec/bibtex/#1.xml}
  {dblp:#1}}
\def\mn@eprint@#1:#2:#3:#4\@nil{\def\@tempa {#1}\def\@tempb {#2}\def\@tempc
  {#3}\ifx \@tempc \@empty \let \@tempc \@tempb \let \@tempb \@tempa \fi \ifx
  \@tempb \@empty \def\@tempb {arXiv}\fi \@ifundefined
  {mn@eprint@\@tempb}{\@tempb:\@tempc}{\expandafter \expandafter \csname
  mn@eprint@\@tempb\endcsname \expandafter{\@tempc}}}

\bibitem[\protect\citeauthoryear{{Abazajian} et~al.,}{{Abazajian}
  et~al.}{2009}]{2009ApJS..182..543A}
{Abazajian} K.~N.,  et~al., 2009, \mn@doi [\apjs]
  {10.1088/0067-0049/182/2/543}, \href
  {http://adsabs.harvard.edu/abs/2009ApJS..182..543A} {182, 543}

\bibitem[\protect\citeauthoryear{{Abbott} et~al.,}{{Abbott}
  et~al.}{2016}]{2016LRR....19....1A}
{Abbott} B.~P.,  et~al., 2016, \mn@doi [Living Reviews in Relativity]
  {10.1007/lrr-2016-1}, \href
  {http://adsabs.harvard.edu/abs/2016LRR....19....1A} {19, 1}

\bibitem[\protect\citeauthoryear{Abbott et~al.,}{Abbott
  et~al.}{2017a}]{PhysRevLett.119.161101}
Abbott B.~P.,  et~al., 2017a, \mn@doi [Phys. Rev. Lett.]
  {10.1103/PhysRevLett.119.161101}, 119, 161101

\bibitem[\protect\citeauthoryear{{Abbott} et~al.,}{{Abbott}
  et~al.}{2017b}]{2017ApJ...848L..12A}
{Abbott} B.~P.,  et~al., 2017b, \mn@doi [\apjl] {10.3847/2041-8213/aa91c9},
  \href {http://adsabs.harvard.edu/abs/2017ApJ...848L..12A} {848, L12}

\bibitem[\protect\citeauthoryear{{Abbott} et~al.,}{{Abbott}
  et~al.}{2017c}]{2017ApJ...848L..13A}
{Abbott} B.~P.,  et~al., 2017c, \mn@doi [\apjl] {10.3847/2041-8213/aa920c},
  \href {http://adsabs.harvard.edu/abs/2017ApJ...848L..13A} {848, L13}

\bibitem[\protect\citeauthoryear{{Acernese} et~al.,}{{Acernese}
  et~al.}{2015}]{2015CQGra..32b4001A}
{Acernese} F.,  et~al., 2015, \mn@doi [Classical and Quantum Gravity]
  {10.1088/0264-9381/32/2/024001}, \href
  {http://adsabs.harvard.edu/abs/2015CQGra..32b4001A} {32, 024001}

\bibitem[\protect\citeauthoryear{{Ackermann} et~al.,}{{Ackermann}
  et~al.}{2015}]{2015ApJ...807..169A}
{Ackermann} M.,  et~al., 2015, \mn@doi [\apj] {10.1088/0004-637X/807/2/169},
  \href {http://adsabs.harvard.edu/abs/2015ApJ...807..169A} {807, 169}

\bibitem[\protect\citeauthoryear{{Adams} et~al.,}{{Adams}
  et~al.}{2018}]{2018PASP..130c4202A}
{Adams} S.~M.,  et~al., 2018, \mn@doi [\pasp] {10.1088/1538-3873/aaa356}, \href
  {http://adsabs.harvard.edu/abs/2018PASP..130c4202A} {130, 034202}

\bibitem[\protect\citeauthoryear{{Andreoni} et~al.,}{{Andreoni}
  et~al.}{2017}]{2017PASA...34...69A}
{Andreoni} I.,  et~al., 2017, \mn@doi [\pasa] {10.1017/pasa.2017.65}, \href
  {http://adsabs.harvard.edu/abs/2017PASA...34...69A} {34, e069}

\bibitem[\protect\citeauthoryear{{Arcavi} et~al.,}{{Arcavi}
  et~al.}{2010}]{2010ATel.3027....1A}
{Arcavi} I.,  et~al., 2010, The Astronomer's Telegram, \href
  {http://adsabs.harvard.edu/abs/2010ATel.3027....1A} {3027}

\bibitem[\protect\citeauthoryear{{Arcavi} et~al.,}{{Arcavi}
  et~al.}{2017}]{2017Natur.551...64A}
{Arcavi} I.,  et~al., 2017, \mn@doi [\nat] {10.1038/nature24291}, \href
  {http://adsabs.harvard.edu/abs/2017Natur.551...64A} {551, 64}

\bibitem[\protect\citeauthoryear{{Astropy Collaboration} et~al.,}{{Astropy
  Collaboration} et~al.}{2013}]{2013A&A...558A..33A}
{Astropy Collaboration} et~al., 2013, \mn@doi [\aap]
  {10.1051/0004-6361/201322068}, \href
  {http://adsabs.harvard.edu/abs/2013A%26A...558A..33A} {558, A33}

\bibitem[\protect\citeauthoryear{Barbary}{Barbary}{2014}]{Barbary:11938}
Barbary K.,  2014, ] {{10.5281/zenodo.11938}}

\bibitem[\protect\citeauthoryear{{Barnes} \& {Kasen}}{{Barnes} \&
  {Kasen}}{2013}]{2013ApJ...775...18B}
{Barnes} J.,  {Kasen} D.,  2013, \mn@doi [\apj] {10.1088/0004-637X/775/1/18},
  \href {http://adsabs.harvard.edu/abs/2013ApJ...775...18B} {775, 18}

\bibitem[\protect\citeauthoryear{{Becker} et~al.,}{{Becker}
  et~al.}{2004}]{2004ApJ...611..418B}
{Becker} A.~C.,  et~al., 2004, \mn@doi [\apj] {10.1086/421994}, \href
  {http://adsabs.harvard.edu/abs/2004ApJ...611..418B} {611, 418}

\bibitem[\protect\citeauthoryear{{Benn}, {Dee}  \& {Ag{\'o}cs}}{{Benn}
  et~al.}{2008}]{2008SPIE.7014E..6XB}
{Benn} C.,  {Dee} K.,   {Ag{\'o}cs} T.,  2008, in Ground-based and Airborne
  Instrumentation for Astronomy II. p. 70146X, \mn@doi{10.1117/12.788694}

\bibitem[\protect\citeauthoryear{{Berger} et~al.,}{{Berger}
  et~al.}{2013}]{2013ApJ...779...18B}
{Berger} E.,  et~al., 2013, \mn@doi [\apj] {10.1088/0004-637X/779/1/18}, \href
  {http://adsabs.harvard.edu/abs/2013ApJ...779...18B} {779, 18}

\bibitem[\protect\citeauthoryear{{Bertin} \& {Arnouts}}{{Bertin} \&
  {Arnouts}}{1996}]{1996A&AS..117..393B}
{Bertin} E.,  {Arnouts} S.,  1996, \mn@doi [\aaps] {10.1051/aas:1996164}, \href
  {http://adsabs.harvard.edu/abs/1996A%26AS..117..393B} {117, 393}

\bibitem[\protect\citeauthoryear{{Best} et~al.,}{{Best}
  et~al.}{2018}]{2018ApJS..234....1B}
{Best} W.~M.~J.,  et~al., 2018, \mn@doi [\apjs] {10.3847/1538-4365/aa9982},
  \href {http://adsabs.harvard.edu/abs/2018ApJS..234....1B} {234, 1}

\bibitem[\protect\citeauthoryear{{Blondin} \& {Tonry}}{{Blondin} \&
  {Tonry}}{2007}]{2007ApJ...666.1024B}
{Blondin} S.,  {Tonry} J.~L.,  2007, \mn@doi [\apj] {10.1086/520494}, \href
  {http://adsabs.harvard.edu/abs/2007ApJ...666.1024B} {666, 1024}

\bibitem[\protect\citeauthoryear{{Bloom} et~al.,}{{Bloom}
  et~al.}{2012}]{2012PASP..124.1175B}
{Bloom} J.~S.,  et~al., 2012, \mn@doi [\pasp] {10.1086/668468}, \href
  {http://adsabs.harvard.edu/abs/2012PASP..124.1175B} {124, 1175}

\bibitem[\protect\citeauthoryear{{Boksenberg}}{{Boksenberg}}{1985}]{1985VA.....28..531B}
{Boksenberg} A.,  1985, \mn@doi [Vistas in Astronomy]
  {10.1016/0083-6656(85)90074-1}, \href
  {http://adsabs.harvard.edu/abs/1985VA.....28..531B} {28, 531}

\bibitem[\protect\citeauthoryear{{Brink}, {Richards}, {Poznanski}, {Bloom},
  {Rice}, {Negahban}  \& {Wainwright}}{{Brink}
  et~al.}{2013}]{2013MNRAS.435.1047B}
{Brink} H.,  {Richards} J.~W.,  {Poznanski} D.,  {Bloom} J.~S.,  {Rice} J.,
  {Negahban} S.,   {Wainwright} M.,  2013, \mn@doi [\mnras]
  {10.1093/mnras/stt1306}, \href
  {http://adsabs.harvard.edu/abs/2013MNRAS.435.1047B} {435, 1047}

\bibitem[\protect\citeauthoryear{{Cao}, {Nugent}  \& {Kasliwal}}{{Cao}
  et~al.}{2016}]{2016PASP..128k4502C}
{Cao} Y.,  {Nugent} P.~E.,   {Kasliwal} M.~M.,  2016, \mn@doi [\pasp]
  {10.1088/1538-3873/128/969/114502}, \href
  {http://adsabs.harvard.edu/abs/2016PASP..128k4502C} {128, 114502}

\bibitem[\protect\citeauthoryear{{Cenko} et~al.,}{{Cenko}
  et~al.}{2013}]{2013ApJ...769..130C}
{Cenko} S.~B.,  et~al., 2013, \mn@doi [\apj] {10.1088/0004-637X/769/2/130},
  \href {http://adsabs.harvard.edu/abs/2013ApJ...769..130C} {769, 130}

\bibitem[\protect\citeauthoryear{{Cenko} et~al.,}{{Cenko}
  et~al.}{2015}]{2015ApJ...803L..24C}
{Cenko} S.~B.,  et~al., 2015, \mn@doi [\apjl] {10.1088/2041-8205/803/2/L24},
  \href {http://adsabs.harvard.edu/abs/2015ApJ...803L..24C} {803, L24}

\bibitem[\protect\citeauthoryear{{Chambers} et~al.,}{{Chambers}
  et~al.}{2016}]{2016arXiv161205560C}
{Chambers} K.~C.,  et~al., 2016, preprint, \href
  {http://adsabs.harvard.edu/abs/2016arXiv161205560C} {} (\mn@eprint {arXiv}
  {1612.05560})

\bibitem[\protect\citeauthoryear{{Condon}, {Cotton}, {Greisen}, {Yin},
  {Perley}, {Taylor}  \& {Broderick}}{{Condon}
  et~al.}{1998}]{1998AJ....115.1693C}
{Condon} J.~J.,  {Cotton} W.~D.,  {Greisen} E.~W.,  {Yin} Q.~F.,  {Perley}
  R.~A.,  {Taylor} G.~B.,   {Broderick} J.~J.,  1998, \mn@doi [\aj]
  {10.1086/300337}, \href {http://adsabs.harvard.edu/abs/1998AJ....115.1693C}
  {115, 1693}

\bibitem[\protect\citeauthoryear{{Coppejans} et~al.,}{{Coppejans}
  et~al.}{2014}]{2014MNRAS.437..510C}
{Coppejans} D.~L.,  et~al., 2014, \mn@doi [\mnras] {10.1093/mnras/stt1900},
  \href {http://adsabs.harvard.edu/abs/2014MNRAS.437..510C} {437, 510}

\bibitem[\protect\citeauthoryear{{Coulter} et~al.,}{{Coulter}
  et~al.}{2017}]{2017Sci...358.1556C}
{Coulter} D.~A.,  et~al., 2017, \mn@doi [Science] {10.1126/science.aap9811},
  \href {http://adsabs.harvard.edu/abs/2017Sci...358.1556C} {358, 1556}

\bibitem[\protect\citeauthoryear{{Cowperthwaite} et~al.,}{{Cowperthwaite}
  et~al.}{2017a}]{2017arXiv171002144C}
{Cowperthwaite} P.~S.,  et~al., 2017a, preprint, \href
  {http://adsabs.harvard.edu/abs/2017arXiv171002144C} {} (\mn@eprint {arXiv}
  {1710.02144})

\bibitem[\protect\citeauthoryear{{Cowperthwaite} et~al.,}{{Cowperthwaite}
  et~al.}{2017b}]{2017ApJ...848L..17C}
{Cowperthwaite} P.~S.,  et~al., 2017b, \mn@doi [\apjl]
  {10.3847/2041-8213/aa8fc7}, \href
  {http://adsabs.harvard.edu/abs/2017ApJ...848L..17C} {848, L17}

\bibitem[\protect\citeauthoryear{{D'Abrusco}, {Massaro}, {Paggi}, {Smith},
  {Masetti}, {Landoni}  \& {Tosti}}{{D'Abrusco}
  et~al.}{2014}]{2014ApJS..215...14D}
{D'Abrusco} R.,  {Massaro} F.,  {Paggi} A.,  {Smith} H.~A.,  {Masetti} N.,
  {Landoni} M.,   {Tosti} G.,  2014, \mn@doi [\apjs]
  {10.1088/0067-0049/215/1/14}, \href
  {http://adsabs.harvard.edu/abs/2014ApJS..215...14D} {215, 14}

\bibitem[\protect\citeauthoryear{{Davenport}, {Becker}, {Kowalski}, {Hawley},
  {Schmidt}, {Hilton}, {Sesar}  \& {Cutri}}{{Davenport}
  et~al.}{2012}]{2012ApJ...748...58D}
{Davenport} J.~R.~A.,  {Becker} A.~C.,  {Kowalski} A.~F.,  {Hawley} S.~L.,
  {Schmidt} S.~J.,  {Hilton} E.~J.,  {Sesar} B.,   {Cutri} R.,  2012, \mn@doi
  [\apj] {10.1088/0004-637X/748/1/58}, \href
  {http://adsabs.harvard.edu/abs/2012ApJ...748...58D} {748, 58}

\bibitem[\protect\citeauthoryear{{D{\'{\i}}az} et~al.,}{{D{\'{\i}}az}
  et~al.}{2017}]{2017ApJ...848L..29D}
{D{\'{\i}}az} M.~C.,  et~al., 2017, \mn@doi [\apjl] {10.3847/2041-8213/aa9060},
  \href {http://adsabs.harvard.edu/abs/2017ApJ...848L..29D} {848, L29}

\bibitem[\protect\citeauthoryear{{Drake} et~al.,}{{Drake}
  et~al.}{2009}]{2009ApJ...696..870D}
{Drake} A.~J.,  et~al., 2009, \mn@doi [\apj] {10.1088/0004-637X/696/1/870},
  \href {http://adsabs.harvard.edu/abs/2009ApJ...696..870D} {696, 870}

\bibitem[\protect\citeauthoryear{{Drake} et~al.,}{{Drake}
  et~al.}{2010a}]{2010CBET.2601....1D}
{Drake} A.~J.,  et~al., 2010a, Central Bureau Electronic Telegrams, \href
  {http://adsabs.harvard.edu/abs/2010CBET.2601....1D} {2601}

\bibitem[\protect\citeauthoryear{{Drake} et~al.,}{{Drake}
  et~al.}{2010b}]{2010ATel.3081....1D}
{Drake} A.~J.,  et~al., 2010b, The Astronomer's Telegram, \href
  {http://adsabs.harvard.edu/abs/2010ATel.3081....1D} {3081}

\bibitem[\protect\citeauthoryear{{Drake} et~al.,}{{Drake}
  et~al.}{2014}]{2014MNRAS.441.1186D}
{Drake} A.~J.,  et~al., 2014, \mn@doi [\mnras] {10.1093/mnras/stu639}, \href
  {http://adsabs.harvard.edu/abs/2014MNRAS.441.1186D} {441, 1186}

\bibitem[\protect\citeauthoryear{{Drout} et~al.,}{{Drout}
  et~al.}{2017}]{2017Sci...358.1570D}
{Drout} M.~R.,  et~al., 2017, \mn@doi [Science] {10.1126/science.aaq0049},
  \href {http://adsabs.harvard.edu/abs/2017Sci...358.1570D} {358, 1570}

\bibitem[\protect\citeauthoryear{{Evans} et~al.,}{{Evans}
  et~al.}{2017}]{2017Sci...358.1565E}
{Evans} P.~A.,  et~al., 2017, \mn@doi [Science] {10.1126/science.aap9580},
  \href {http://adsabs.harvard.edu/abs/2017Sci...358.1565E} {358, 1565}

\bibitem[\protect\citeauthoryear{{Filippenko}}{{Filippenko}}{1997}]{1997ARA&A..35..309F}
{Filippenko} A.~V.,  1997, \mn@doi [\araa] {10.1146/annurev.astro.35.1.309},
  \href {http://adsabs.harvard.edu/abs/1997ARA%26A..35..309F} {35, 309}

\bibitem[\protect\citeauthoryear{{Filippenko} et~al.,}{{Filippenko}
  et~al.}{1992a}]{1992AJ....104.1543F}
{Filippenko} A.~V.,  et~al., 1992a, \mn@doi [\aj] {10.1086/116339}, \href
  {http://adsabs.harvard.edu/abs/1992AJ....104.1543F} {104, 1543}

\bibitem[\protect\citeauthoryear{{Filippenko} et~al.,}{{Filippenko}
  et~al.}{1992b}]{1992ApJ...384L..15F}
{Filippenko} A.~V.,  et~al., 1992b, \mn@doi [\apjl] {10.1086/186252}, \href
  {http://adsabs.harvard.edu/abs/1992ApJ...384L..15F} {384, L15}

\bibitem[\protect\citeauthoryear{{Flewelling} et~al.,}{{Flewelling}
  et~al.}{2016}]{2016arXiv161205243F}
{Flewelling} H.~A.,  et~al., 2016, preprint, \href
  {http://adsabs.harvard.edu/abs/2016arXiv161205243F} {} (\mn@eprint {arXiv}
  {1612.05243})

\bibitem[\protect\citeauthoryear{{Fong}, {Berger}, {Margutti}  \&
  {Zauderer}}{{Fong} et~al.}{2015}]{2015ApJ...815..102F}
{Fong} W.,  {Berger} E.,  {Margutti} R.,   {Zauderer} B.~A.,  2015, \mn@doi
  [\apj] {10.1088/0004-637X/815/2/102}, \href
  {http://adsabs.harvard.edu/abs/2015ApJ...815..102F} {815, 102}

\bibitem[\protect\citeauthoryear{{Frohmaier}, {Sullivan}, {Nugent}, {Goldstein}
   \& {DeRose}}{{Frohmaier} et~al.}{2017}]{2017ApJS..230....4F}
{Frohmaier} C.,  {Sullivan} M.,  {Nugent} P.~E.,  {Goldstein} D.~A.,   {DeRose}
  J.,  2017, \mn@doi [\apjs] {10.3847/1538-4365/aa6d70}, \href
  {http://adsabs.harvard.edu/abs/2017ApJS..230....4F} {230, 4}

\bibitem[\protect\citeauthoryear{{Gehrels}}{{Gehrels}}{1986}]{Gehrels:1986}
{Gehrels} N.,  1986, \mn@doi [\apj] {10.1086/164079}, \href
  {http://adsabs.harvard.edu/abs/1986ApJ...303..336G} {303, 336}

\bibitem[\protect\citeauthoryear{{Gehrels}, {Cannizzo}, {Kanner}, {Kasliwal},
  {Nissanke}  \& {Singer}}{{Gehrels} et~al.}{2016}]{2016ApJ...820..136G}
{Gehrels} N.,  {Cannizzo} J.~K.,  {Kanner} J.,  {Kasliwal} M.~M.,  {Nissanke}
  S.,   {Singer} L.~P.,  2016, \mn@doi [\apj] {10.3847/0004-637X/820/2/136},
  \href {http://adsabs.harvard.edu/abs/2016ApJ...820..136G} {820, 136}

\bibitem[\protect\citeauthoryear{{Gershberg}}{{Gershberg}}{1972}]{1972Ap&SS..19...75G}
{Gershberg} R.~E.,  1972, \mn@doi [\apss] {10.1007/BF00643168}, \href
  {http://adsabs.harvard.edu/abs/1972Ap%26SS..19...75G} {19, 75}

\bibitem[\protect\citeauthoryear{{Gezari} et~al.,}{{Gezari}
  et~al.}{2017}]{2017ApJ...835..144G}
{Gezari} S.,  et~al., 2017, \mn@doi [\apj] {10.3847/1538-4357/835/2/144}, \href
  {https://ui.adsabs.harvard.edu/\#abs/2017ApJ...835..144G} {835, 144}

\bibitem[\protect\citeauthoryear{{Gieseke} et~al.,}{{Gieseke}
  et~al.}{2017}]{2017MNRAS.472.3101G}
{Gieseke} F.,  et~al., 2017, \mn@doi [\mnras] {10.1093/mnras/stx2161}, \href
  {http://adsabs.harvard.edu/abs/2017MNRAS.472.3101G} {472, 3101}

\bibitem[\protect\citeauthoryear{{Gilliland}, {Nugent}  \&
  {Phillips}}{{Gilliland} et~al.}{1999}]{1999ApJ...521...30G}
{Gilliland} R.~L.,  {Nugent} P.~E.,   {Phillips} M.~M.,  1999, \mn@doi [\apj]
  {10.1086/307549}, \href {http://adsabs.harvard.edu/abs/1999ApJ...521...30G}
  {521, 30}

\bibitem[\protect\citeauthoryear{{Graur} et~al.,}{{Graur}
  et~al.}{2014}]{2014ApJ...783...28G}
{Graur} O.,  et~al., 2014, \mn@doi [\apj] {10.1088/0004-637X/783/1/28}, \href
  {http://adsabs.harvard.edu/abs/2014ApJ...783...28G} {783, 28}

\bibitem[\protect\citeauthoryear{{Guy} et~al.,}{{Guy}
  et~al.}{2007}]{2007A&A...466...11G}
{Guy} J.,  et~al., 2007, \mn@doi [\aap] {10.1051/0004-6361:20066930}, \href
  {http://adsabs.harvard.edu/abs/2007A%26A...466...11G} {466, 11}

\bibitem[\protect\citeauthoryear{{Harrison}, {Johnson}, {McArthur}, {Benedict},
  {Szkody}, {Howell}  \& {Gelino}}{{Harrison}
  et~al.}{2004}]{2004AJ....127..460H}
{Harrison} T.~E.,  {Johnson} J.~J.,  {McArthur} B.~E.,  {Benedict} G.~F.,
  {Szkody} P.,  {Howell} S.~B.,   {Gelino} D.~M.,  2004, \mn@doi [\aj]
  {10.1086/380228}, \href {http://adsabs.harvard.edu/abs/2004AJ....127..460H}
  {127, 460}

\bibitem[\protect\citeauthoryear{{Hawley}, {Davenport}, {Kowalski},
  {Wisniewski}, {Hebb}, {Deitrick}  \& {Hilton}}{{Hawley}
  et~al.}{2014}]{2014ApJ...797..121H}
{Hawley} S.~L.,  {Davenport} J.~R.~A.,  {Kowalski} A.~F.,  {Wisniewski} J.~P.,
  {Hebb} L.,  {Deitrick} R.,   {Hilton} E.~J.,  2014, \mn@doi [\apj]
  {10.1088/0004-637X/797/2/121}, \href
  {http://adsabs.harvard.edu/abs/2014ApJ...797..121H} {797, 121}

\bibitem[\protect\citeauthoryear{{Hilton}}{{Hilton}}{2011}]{2011PhDT.......144H}
{Hilton} E.~J.,  2011, PhD thesis, University of Washington

\bibitem[\protect\citeauthoryear{{Hilton}, {West}, {Hawley}  \&
  {Kowalski}}{{Hilton} et~al.}{2010}]{2010AJ....140.1402H}
{Hilton} E.~J.,  {West} A.~A.,  {Hawley} S.~L.,   {Kowalski} A.~F.,  2010,
  \mn@doi [\aj] {10.1088/0004-6256/140/5/1402}, \href
  {http://adsabs.harvard.edu/abs/2010AJ....140.1402H} {140, 1402}

\bibitem[\protect\citeauthoryear{Ho et~al.,}{Ho et~al.}{2018}]{Ho2018}
Ho A. Y.~Q.,  et~al., 2018, The Astrophysical Journal Letters, 854, L13

\bibitem[\protect\citeauthoryear{{Hsiao}, {Conley}, {Howell}, {Sullivan},
  {Pritchet}, {Carlberg}, {Nugent}  \& {Phillips}}{{Hsiao}
  et~al.}{2007}]{2007ApJ...663.1187H}
{Hsiao} E.~Y.,  {Conley} A.,  {Howell} D.~A.,  {Sullivan} M.,  {Pritchet}
  C.~J.,  {Carlberg} R.~G.,  {Nugent} P.~E.,   {Phillips} M.~M.,  2007, \mn@doi
  [\apj] {10.1086/518232}, \href
  {http://adsabs.harvard.edu/abs/2007ApJ...663.1187H} {663, 1187}

\bibitem[\protect\citeauthoryear{{Hu} et~al.,}{{Hu}
  et~al.}{2017}]{2017SciBu..62.1433H}
{Hu} L.,  et~al., 2017, \mn@doi [Science Bulletin, Vol.~62, No.21, p.1433-1438,
  2017] {10.1016/j.scib.2017.10.006}, \href
  {http://adsabs.harvard.edu/abs/2017SciBu..62.1433H} {62, 1433}

\bibitem[\protect\citeauthoryear{{Huchra} et~al.,}{{Huchra}
  et~al.}{2012}]{2012ApJS..199...26H}
{Huchra} J.~P.,  et~al., 2012, \mn@doi [The Astrophysical Journal Supplement
  Series] {10.1088/0067-0049/199/2/26}, \href
  {https://ui.adsabs.harvard.edu/#abs/2012ApJS..199...26H} {199, 26}

\bibitem[\protect\citeauthoryear{Hunter}{Hunter}{2007}]{Hunter:2007}
Hunter J.~D.,  2007, Computing In Science \& Engineering, 9, 90

\bibitem[\protect\citeauthoryear{{Kasen}, {Fern{\'a}ndez}  \&
  {Metzger}}{{Kasen} et~al.}{2015}]{2015MNRAS.450.1777K}
{Kasen} D.,  {Fern{\'a}ndez} R.,   {Metzger} B.~D.,  2015, \mn@doi [\mnras]
  {10.1093/mnras/stv721}, \href
  {http://adsabs.harvard.edu/abs/2015MNRAS.450.1777K} {450, 1777}

\bibitem[\protect\citeauthoryear{{Kasen}, {Metzger}, {Barnes}, {Quataert}  \&
  {Ramirez-Ruiz}}{{Kasen} et~al.}{2017}]{2017Natur.551...80K}
{Kasen} D.,  {Metzger} B.,  {Barnes} J.,  {Quataert} E.,   {Ramirez-Ruiz} E.,
  2017, \mn@doi [\nat] {10.1038/nature24453}, \href
  {http://adsabs.harvard.edu/abs/2017Natur.551...80K} {551, 80}

\bibitem[\protect\citeauthoryear{{Kasliwal}}{{Kasliwal}}{2011}]{2011PhDT........35K}
{Kasliwal} M.~M.,  2011, PhD thesis, California Institute of Technology

\bibitem[\protect\citeauthoryear{{Kasliwal} et~al.,}{{Kasliwal}
  et~al.}{2017}]{2017Sci...358.1559K}
{Kasliwal} M.~M.,  et~al., 2017, \mn@doi [Science] {10.1126/science.aap9455},
  \href {http://adsabs.harvard.edu/abs/2017Sci...358.1559K} {358, 1559}

\bibitem[\protect\citeauthoryear{{Kato} et~al.,}{{Kato}
  et~al.}{2009}]{2009PASJ...61S.395K}
{Kato} T.,  et~al., 2009, \mn@doi [\pasj] {10.1093/pasj/61.sp2.S395}, \href
  {http://adsabs.harvard.edu/abs/2009PASJ...61S.395K} {61, S395}

\bibitem[\protect\citeauthoryear{{Kowalski}, {Hawley}, {Hilton}, {Becker},
  {West}, {Bochanski}  \& {Sesar}}{{Kowalski}
  et~al.}{2009}]{2009AJ....138..633K}
{Kowalski} A.~F.,  {Hawley} S.~L.,  {Hilton} E.~J.,  {Becker} A.~C.,  {West}
  A.~A.,  {Bochanski} J.~J.,   {Sesar} B.,  2009, \mn@doi [\aj]
  {10.1088/0004-6256/138/2/633}, \href
  {http://adsabs.harvard.edu/abs/2009AJ....138..633K} {138, 633}

\bibitem[\protect\citeauthoryear{{Kowalski}, {Hawley}, {Holtzman}, {Wisniewski}
   \& {Hilton}}{{Kowalski} et~al.}{2010}]{2010ApJ...714L..98K}
{Kowalski} A.~F.,  {Hawley} S.~L.,  {Holtzman} J.~A.,  {Wisniewski} J.~P.,
  {Hilton} E.~J.,  2010, \mn@doi [\apjl] {10.1088/2041-8205/714/1/L98}, \href
  {http://adsabs.harvard.edu/abs/2010ApJ...714L..98K} {714, L98}

\bibitem[\protect\citeauthoryear{{Kulkarni}}{{Kulkarni}}{2005}]{2005astro.ph.10256K}
{Kulkarni} S.~R.,  2005, ArXiv Astrophysics e-prints, \href
  {http://adsabs.harvard.edu/abs/2005astro.ph.10256K} {}

\bibitem[\protect\citeauthoryear{{Kulkarni} \& {Rau}}{{Kulkarni} \&
  {Rau}}{2006}]{2006ApJ...644L..63K}
{Kulkarni} S.~R.,  {Rau} A.,  2006, \mn@doi [\apjl] {10.1086/505423}, \href
  {http://adsabs.harvard.edu/abs/2006ApJ...644L..63K} {644, L63}

\bibitem[\protect\citeauthoryear{{Kulkarni}, {Perley}  \& {Miller}}{{Kulkarni}
  et~al.}{2018}]{2018ApJ...860...22K}
{Kulkarni} S.~R.,  {Perley} D.~A.,   {Miller} A.~A.,  2018, \mn@doi [\apj]
  {10.3847/1538-4357/aabf85}, \href
  {http://adsabs.harvard.edu/abs/2018ApJ...860...22K} {860, 22}

\bibitem[\protect\citeauthoryear{{LIGO Scientific Collaboration} et~al.,}{{LIGO
  Scientific Collaboration} et~al.}{2015}]{2015CQGra..32g4001L}
{LIGO Scientific Collaboration} et~al., 2015, \mn@doi [Classical and Quantum
  Gravity] {10.1088/0264-9381/32/7/074001}, \href
  {http://adsabs.harvard.edu/abs/2015CQGra..32g4001L} {32, 074001}

\bibitem[\protect\citeauthoryear{{Law} et~al.,}{{Law}
  et~al.}{2009}]{2009PASP..121.1395L}
{Law} N.~M.,  et~al., 2009, \mn@doi [\pasp] {10.1086/648598}, \href
  {http://adsabs.harvard.edu/abs/2009PASP..121.1395L} {121, 1395}

\bibitem[\protect\citeauthoryear{{Li} \& {Paczy{\'n}ski}}{{Li} \&
  {Paczy{\'n}ski}}{1998}]{1998ApJ...507L..59L}
{Li} L.-X.,  {Paczy{\'n}ski} B.,  1998, \mn@doi [\apjl] {10.1086/311680}, \href
  {http://adsabs.harvard.edu/abs/1998ApJ...507L..59L} {507, L59}

\bibitem[\protect\citeauthoryear{{Li} et~al.,}{{Li}
  et~al.}{2011}]{2011MNRAS.412.1441L}
{Li} W.,  et~al., 2011, \mn@doi [\mnras] {10.1111/j.1365-2966.2011.18160.x},
  \href {http://adsabs.harvard.edu/abs/2011MNRAS.412.1441L} {412, 1441}

\bibitem[\protect\citeauthoryear{{Lipunov} et~al.,}{{Lipunov}
  et~al.}{2007}]{2007ARep...51.1004L}
{Lipunov} V.~M.,  et~al., 2007, \mn@doi [Astronomy Reports]
  {10.1134/S1063772907120050}, \href
  {http://adsabs.harvard.edu/abs/2007ARep...51.1004L} {51, 1004}

\bibitem[\protect\citeauthoryear{{Lipunov}, {Kornilov}, {Gorbovskoy},
  {Lipunova}, {Vlasenko}, {Panchenko}, {Tyurina}  \& {Grinshpun}}{{Lipunov}
  et~al.}{2018}]{2018NewA...63...48L}
{Lipunov} V.,  {Kornilov} V.,  {Gorbovskoy} E.,  {Lipunova} G.,  {Vlasenko} D.,
   {Panchenko} I.,  {Tyurina} N.,   {Grinshpun} V.,  2018, \mn@doi [\na]
  {10.1016/j.newast.2018.02.004}, \href
  {http://adsabs.harvard.edu/abs/2018NewA...63...48L} {63, 48}

\bibitem[\protect\citeauthoryear{{Maguire} et~al.,}{{Maguire}
  et~al.}{2012}]{2012MNRAS.426.2359M}
{Maguire} K.,  et~al., 2012, \mn@doi [\mnras]
  {10.1111/j.1365-2966.2012.21909.x}, \href
  {http://adsabs.harvard.edu/abs/2012MNRAS.426.2359M} {426, 2359}

\bibitem[\protect\citeauthoryear{{Mandel}, {Scolnic}, {Shariff}, {Foley}  \&
  {Kirshner}}{{Mandel} et~al.}{2017}]{2017ApJ...842...93M}
{Mandel} K.~S.,  {Scolnic} D.~M.,  {Shariff} H.,  {Foley} R.~J.,   {Kirshner}
  R.~P.,  2017, \mn@doi [\apj] {10.3847/1538-4357/aa6038}, \href
  {http://adsabs.harvard.edu/abs/2017ApJ...842...93M} {842, 93}

\bibitem[\protect\citeauthoryear{{Metzger} \& {Fern{\'a}ndez}}{{Metzger} \&
  {Fern{\'a}ndez}}{2014}]{2014MNRAS.441.3444M}
{Metzger} B.~D.,  {Fern{\'a}ndez} R.,  2014, \mn@doi [\mnras]
  {10.1093/mnras/stu802}, \href
  {http://adsabs.harvard.edu/abs/2014MNRAS.441.3444M} {441, 3444}

\bibitem[\protect\citeauthoryear{{Metzger}, {Arcones}, {Quataert}  \&
  {Mart{\'{\i}}nez-Pinedo}}{{Metzger} et~al.}{2010}]{2010MNRAS.402.2771M}
{Metzger} B.~D.,  {Arcones} A.,  {Quataert} E.,   {Mart{\'{\i}}nez-Pinedo} G.,
  2010, \mn@doi [\mnras] {10.1111/j.1365-2966.2009.16107.x}, \href
  {http://adsabs.harvard.edu/abs/2010MNRAS.402.2771M} {402, 2771}

\bibitem[\protect\citeauthoryear{{Mickaelian}, {Hovhannisyan}, {Engels},
  {Hagen}  \& {Voges}}{{Mickaelian} et~al.}{2006}]{2006A&A...449..425M}
{Mickaelian} A.~M.,  {Hovhannisyan} L.~R.,  {Engels} D.,  {Hagen} H.-J.,
  {Voges} W.,  2006, \mn@doi [\aap] {10.1051/0004-6361:20053967}, \href
  {http://adsabs.harvard.edu/abs/2006A%26A...449..425M} {449, 425}

\bibitem[\protect\citeauthoryear{{Miller}, {Kulkarni}, {Cao}, {Laher}, {Masci}
  \& {Surace}}{{Miller} et~al.}{2017}]{2017AJ....153...73M}
{Miller} A.~A.,  {Kulkarni} M.~K.,  {Cao} Y.,  {Laher} R.~R.,  {Masci} F.~J.,
  {Surace} J.~A.,  2017, \mn@doi [\aj] {10.3847/1538-3881/153/2/73}, \href
  {http://adsabs.harvard.edu/abs/2017AJ....153...73M} {153, 73}

\bibitem[\protect\citeauthoryear{{Nelemans}, {Yungelson}  \& {Portegies
  Zwart}}{{Nelemans} et~al.}{2004}]{2004MNRAS.349..181N}
{Nelemans} G.,  {Yungelson} L.~R.,   {Portegies Zwart} S.~F.,  2004, \mn@doi
  [\mnras] {10.1111/j.1365-2966.2004.07479.x}, \href
  {http://adsabs.harvard.edu/abs/2004MNRAS.349..181N} {349, 181}

\bibitem[\protect\citeauthoryear{{Oke} \& {Gunn}}{{Oke} \&
  {Gunn}}{1982}]{1982PASP...94..586O}
{Oke} J.~B.,  {Gunn} J.~E.,  1982, \mn@doi [\pasp] {10.1086/131027}, \href
  {http://adsabs.harvard.edu/abs/1982PASP...94..586O} {94, 586}

\bibitem[\protect\citeauthoryear{{Oke} et~al.,}{{Oke}
  et~al.}{1995}]{1995PASP..107..375O}
{Oke} J.~B.,  et~al., 1995, \mn@doi [\pasp] {10.1086/133562}, \href
  {http://adsabs.harvard.edu/abs/1995PASP..107..375O} {107, 375}

\bibitem[\protect\citeauthoryear{{Oliveira}, {Rodrigues}, {Cieslinski},
  {Jablonski}, {Silva}, {Almeida}, {Rodr{\'{\i}}guez-Ardila}  \&
  {Palhares}}{{Oliveira} et~al.}{2017}]{2017AJ....153..144O}
{Oliveira} A.~S.,  {Rodrigues} C.~V.,  {Cieslinski} D.,  {Jablonski} F.~J.,
  {Silva} K.~M.~G.,  {Almeida} L.~A.,  {Rodr{\'{\i}}guez-Ardila} A.,
  {Palhares} M.~S.,  2017, \mn@doi [\aj] {10.3847/1538-3881/aa610d}, \href
  {http://adsabs.harvard.edu/abs/2017AJ....153..144O} {153, 144}

\bibitem[\protect\citeauthoryear{{Osaki} \& {Kato}}{{Osaki} \&
  {Kato}}{2013}]{2013PASJ...65...50O}
{Osaki} Y.,  {Kato} T.,  2013, \mn@doi [\pasj] {10.1093/pasj/65.3.50}, \href
  {http://adsabs.harvard.edu/abs/2013PASJ...65...50O} {65, 50}

\bibitem[\protect\citeauthoryear{{Pan} et~al.,}{{Pan}
  et~al.}{2014}]{2014MNRAS.438.1391P}
{Pan} Y.-C.,  et~al., 2014, \mn@doi [\mnras] {10.1093/mnras/stt2287}, \href
  {http://adsabs.harvard.edu/abs/2014MNRAS.438.1391P} {438, 1391}

\bibitem[\protect\citeauthoryear{{Paturel}, {Petit}, {Prugniel}, {Theureau},
  {Rousseau}, {Brouty}, {Dubois}  \& {Cambr{\'e}sy}}{{Paturel}
  et~al.}{2003}]{2003A&A...412...45P}
{Paturel} G.,  {Petit} C.,  {Prugniel} P.,  {Theureau} G.,  {Rousseau} J.,
  {Brouty} M.,  {Dubois} P.,   {Cambr{\'e}sy} L.,  2003, \mn@doi [\aap]
  {10.1051/0004-6361:20031411}, \href
  {http://adsabs.harvard.edu/abs/2003A%26A...412...45P} {412, 45}

\bibitem[\protect\citeauthoryear{{Pecaut} \& {Mamajek}}{{Pecaut} \&
  {Mamajek}}{2013}]{2013ApJS..208....9P}
{Pecaut} M.~J.,  {Mamajek} E.~E.,  2013, \mn@doi [\apjs]
  {10.1088/0067-0049/208/1/9}, \href
  {http://adsabs.harvard.edu/abs/2013ApJS..208....9P} {208, 9}

\bibitem[\protect\citeauthoryear{{Pian} et~al.,}{{Pian}
  et~al.}{2017}]{2017Natur.551...67P}
{Pian} E.,  et~al., 2017, \mn@doi [\nat] {10.1038/nature24298}, \href
  {http://adsabs.harvard.edu/abs/2017Natur.551...67P} {551, 67}

\bibitem[\protect\citeauthoryear{{Piran}}{{Piran}}{1999}]{1999PhR...314..575P}
{Piran} T.,  1999, \mn@doi [\physrep] {10.1016/S0370-1573(98)00127-6}, \href
  {http://adsabs.harvard.edu/abs/1999PhR...314..575P} {314, 575}

\bibitem[\protect\citeauthoryear{{Pozanenko} et~al.,}{{Pozanenko}
  et~al.}{2018}]{2018ApJ...852L..30P}
{Pozanenko} A.~S.,  et~al., 2018, \mn@doi [\apjl] {10.3847/2041-8213/aaa2f6},
  \href {http://adsabs.harvard.edu/abs/2018ApJ...852L..30P} {852, L30}

\bibitem[\protect\citeauthoryear{{Pretorius} \& {Knigge}}{{Pretorius} \&
  {Knigge}}{2012}]{2012MNRAS.419.1442P}
{Pretorius} M.~L.,  {Knigge} C.,  2012, \mn@doi [\mnras]
  {10.1111/j.1365-2966.2011.19801.x}, \href
  {http://adsabs.harvard.edu/abs/2012MNRAS.419.1442P} {419, 1442}

\bibitem[\protect\citeauthoryear{{Rau}, {Ofek}, {Kulkarni}, {Madore},
  {Pevunova}  \& {Ajello}}{{Rau} et~al.}{2008}]{2008ApJ...682.1205R}
{Rau} A.,  {Ofek} E.~O.,  {Kulkarni} S.~R.,  {Madore} B.~F.,  {Pevunova} O.,
  {Ajello} M.,  2008, \mn@doi [\apj] {10.1086/589762}, \href
  {http://adsabs.harvard.edu/abs/2008ApJ...682.1205R} {682, 1205}

\bibitem[\protect\citeauthoryear{{Rau} et~al.,}{{Rau}
  et~al.}{2009}]{2009PASP..121.1334R}
{Rau} A.,  et~al., 2009, \mn@doi [\pasp] {10.1086/605911}, \href
  {http://adsabs.harvard.edu/abs/2009PASP..121.1334R} {121, 1334}

\bibitem[\protect\citeauthoryear{{Roberts}, {Kasen}, {Lee}  \&
  {Ramirez-Ruiz}}{{Roberts} et~al.}{2011}]{2011ApJ...736L..21R}
{Roberts} L.~F.,  {Kasen} D.,  {Lee} W.~H.,   {Ramirez-Ruiz} E.,  2011, \mn@doi
  [\apjl] {10.1088/2041-8205/736/1/L21}, \href
  {http://adsabs.harvard.edu/abs/2011ApJ...736L..21R} {736, L21}

\bibitem[\protect\citeauthoryear{{Rosswog}, {Feindt}, {Korobkin}, {Wu},
  {Sollerman}, {Goobar}  \& {Martinez-Pinedo}}{{Rosswog}
  et~al.}{2017}]{2017CQGra..34j4001R}
{Rosswog} S.,  {Feindt} U.,  {Korobkin} O.,  {Wu} M.-R.,  {Sollerman} J.,
  {Goobar} A.,   {Martinez-Pinedo} G.,  2017, \mn@doi [Classical and Quantum
  Gravity] {10.1088/1361-6382/aa68a9}, \href
  {http://adsabs.harvard.edu/abs/2017CQGra..34j4001R} {34, 104001}

\bibitem[\protect\citeauthoryear{{Rykoff} et~al.,}{{Rykoff}
  et~al.}{2005}]{2005ApJ...631.1032R}
{Rykoff} E.~S.,  et~al., 2005, \mn@doi [\apj] {10.1086/432832}, \href
  {http://adsabs.harvard.edu/abs/2005ApJ...631.1032R} {631, 1032}

\bibitem[\protect\citeauthoryear{{Schlegel}, {Finkbeiner}  \&
  {Davis}}{{Schlegel} et~al.}{1998}]{1998ApJ...500..525S}
{Schlegel} D.~J.,  {Finkbeiner} D.~P.,   {Davis} M.,  1998, \mn@doi [\apj]
  {10.1086/305772}, \href {http://adsabs.harvard.edu/abs/1998ApJ...500..525S}
  {500, 525}

\bibitem[\protect\citeauthoryear{{Schmidt} et~al.,}{{Schmidt}
  et~al.}{2014}]{2014ApJ...781L..24S}
{Schmidt} S.~J.,  et~al., 2014, \mn@doi [\apjl] {10.1088/2041-8205/781/2/L24},
  \href {http://adsabs.harvard.edu/abs/2014ApJ...781L..24S} {781, L24}

\bibitem[\protect\citeauthoryear{{Shappee} et~al.,}{{Shappee}
  et~al.}{2017}]{2017Sci...358.1574S}
{Shappee} B.~J.,  et~al., 2017, \mn@doi [Science] {10.1126/science.aaq0186},
  \href {http://adsabs.harvard.edu/abs/2017Sci...358.1574S} {358, 1574}

\bibitem[\protect\citeauthoryear{{Silverberg}, {Kowalski}, {Davenport},
  {Wisniewski}, {Hawley}  \& {Hilton}}{{Silverberg}
  et~al.}{2016}]{2016ApJ...829..129S}
{Silverberg} S.~M.,  {Kowalski} A.~F.,  {Davenport} J.~R.~A.,  {Wisniewski}
  J.~P.,  {Hawley} S.~L.,   {Hilton} E.~J.,  2016, \mn@doi [\apj]
  {10.3847/0004-637X/829/2/129}, \href
  {http://adsabs.harvard.edu/abs/2016ApJ...829..129S} {829, 129}

\bibitem[\protect\citeauthoryear{{Singer} et~al.,}{{Singer}
  et~al.}{2015}]{2015ApJ...806...52S}
{Singer} L.~P.,  et~al., 2015, \mn@doi [\apj] {10.1088/0004-637X/806/1/52},
  \href {http://adsabs.harvard.edu/abs/2015ApJ...806...52S} {806, 52}

\bibitem[\protect\citeauthoryear{{Smartt} et~al.,}{{Smartt}
  et~al.}{2017}]{2017Natur.551...75S}
{Smartt} S.~J.,  et~al., 2017, \mn@doi [\nat] {10.1038/nature24303}, \href
  {http://adsabs.harvard.edu/abs/2017Natur.551...75S} {551, 75}

\bibitem[\protect\citeauthoryear{{Smith} et~al.,}{{Smith}
  et~al.}{2011}]{2011MNRAS.412.1309S}
{Smith} A.~M.,  et~al., 2011, \mn@doi [\mnras]
  {10.1111/j.1365-2966.2010.17994.x}, \href
  {http://adsabs.harvard.edu/abs/2011MNRAS.412.1309S} {412, 1309}

\bibitem[\protect\citeauthoryear{{Szkody}, {Everett}, {Howell}, {Landolt},
  {Bond}, {Silva}  \& {Vasquez-Soltero}}{{Szkody}
  et~al.}{2014}]{2014AJ....148...63S}
{Szkody} P.,  {Everett} M.~E.,  {Howell} S.~B.,  {Landolt} A.~U.,  {Bond}
  H.~E.,  {Silva} D.~R.,   {Vasquez-Soltero} S.,  2014, \mn@doi [\aj]
  {10.1088/0004-6256/148/4/63}, \href
  {http://adsabs.harvard.edu/abs/2014AJ....148...63S} {148, 63}

\bibitem[\protect\citeauthoryear{{Tanaka} \& {Hotokezaka}}{{Tanaka} \&
  {Hotokezaka}}{2013}]{2013ApJ...775..113T}
{Tanaka} M.,  {Hotokezaka} K.,  2013, \mn@doi [\apj]
  {10.1088/0004-637X/775/2/113}, \href
  {http://adsabs.harvard.edu/abs/2013ApJ...775..113T} {775, 113}

\bibitem[\protect\citeauthoryear{{Tanvir} et~al.,}{{Tanvir}
  et~al.}{2017}]{2017ApJ...848L..27T}
{Tanvir} N.~R.,  et~al., 2017, \mn@doi [\apjl] {10.3847/2041-8213/aa90b6},
  \href {http://adsabs.harvard.edu/abs/2017ApJ...848L..27T} {848, L27}

\bibitem[\protect\citeauthoryear{{Taubenberger}}{{Taubenberger}}{2017}]{2017arXiv170300528T}
{Taubenberger} S.,  2017, preprint, \href
  {http://adsabs.harvard.edu/abs/2017arXiv170300528T} {} (\mn@eprint {arXiv}
  {1703.00528})

\bibitem[\protect\citeauthoryear{{Troja} et~al.,}{{Troja}
  et~al.}{2017}]{2017Natur.551...71T}
{Troja} E.,  et~al., 2017, \mn@doi [\nat] {10.1038/nature24290}, \href
  {http://adsabs.harvard.edu/abs/2017Natur.551...71T} {551, 71}

\bibitem[\protect\citeauthoryear{{Utsumi} et~al.,}{{Utsumi}
  et~al.}{2017}]{2017PASJ...69..101U}
{Utsumi} Y.,  et~al., 2017, \mn@doi [\pasj] {10.1093/pasj/psx118}, \href
  {http://adsabs.harvard.edu/abs/2017PASJ...69..101U} {69, 101}

\bibitem[\protect\citeauthoryear{{Utsumi} et~al.,}{{Utsumi}
  et~al.}{2018}]{2018PASJ...70....1U}
{Utsumi} Y.,  et~al., 2018, \mn@doi [\pasj] {10.1093/pasj/psx125}, \href
  {http://adsabs.harvard.edu/abs/2018PASJ...70....1U} {70, 1}

\bibitem[\protect\citeauthoryear{{Valenti} et~al.,}{{Valenti}
  et~al.}{2017}]{2017ApJ...848L..24V}
{Valenti} S.,  et~al., 2017, \mn@doi [\apjl] {10.3847/2041-8213/aa8edf}, \href
  {http://adsabs.harvard.edu/abs/2017ApJ...848L..24V} {848, L24}

\bibitem[\protect\citeauthoryear{{Vida}, {K{\H o}v{\'a}ri}, {P{\'a}l},
  {Ol{\'a}h}  \& {Kriskovics}}{{Vida} et~al.}{2017}]{2017ApJ...841..124V}
{Vida} K.,  {K{\H o}v{\'a}ri} Z.,  {P{\'a}l} A.,  {Ol{\'a}h} K.,   {Kriskovics}
  L.,  2017, \mn@doi [\apj] {10.3847/1538-4357/aa6f05}, \href
  {http://adsabs.harvard.edu/abs/2017ApJ...841..124V} {841, 124}

\bibitem[\protect\citeauthoryear{{Villar} et~al.,}{{Villar}
  et~al.}{2017}]{2017ApJ...851L..21V}
{Villar} V.~A.,  et~al., 2017, \mn@doi [\apjl] {10.3847/2041-8213/aa9c84},
  \href {http://adsabs.harvard.edu/abs/2017ApJ...851L..21V} {851, L21}

\bibitem[\protect\citeauthoryear{{Warner}}{{Warner}}{1987}]{1987MNRAS.227...23W}
{Warner} B.,  1987, \mn@doi [\mnras] {10.1093/mnras/227.1.23}, \href
  {http://adsabs.harvard.edu/abs/1987MNRAS.227...23W} {227, 23}

\bibitem[\protect\citeauthoryear{{Warner}}{{Warner}}{2003}]{2003cvs..book.....W}
{Warner} B.,  2003, {Cataclysmic Variable Stars},
  \mn@doi{10.1017/CB0978O511586491.
}

\bibitem[\protect\citeauthoryear{{Wittman} et~al.,}{{Wittman}
  et~al.}{2002}]{2002SPIE.4836...73W}
{Wittman} D.~M.,  et~al., 2002, in {Tyson} J.~A.,  {Wolff} S.,  eds,  \procspie
  Vol. 4836, Survey and Other Telescope Technologies and Discoveries. pp 73--82
  (\mn@eprint {} {astro-ph/0210118}), \mn@doi{10.1117/12.457348}

\bibitem[\protect\citeauthoryear{{Yaron} \& {Gal-Yam}}{{Yaron} \&
  {Gal-Yam}}{2012}]{2012PASP..124..668Y}
{Yaron} O.,  {Gal-Yam} A.,  2012, \mn@doi [\pasp] {10.1086/666656}, \href
  {http://adsabs.harvard.edu/abs/2012PASP..124..668Y} {124, 668}

\bibitem[\protect\citeauthoryear{{Zackay}, {Ofek}  \& {Gal-Yam}}{{Zackay}
  et~al.}{2016}]{2016ApJ...830...27Z}
{Zackay} B.,  {Ofek} E.~O.,   {Gal-Yam} A.,  2016, \mn@doi [\apj]
  {10.3847/0004-637X/830/1/27}, \href
  {http://adsabs.harvard.edu/abs/2016ApJ...830...27Z} {830, 27}

\makeatother
\end{thebibliography}



\appendix

\section{Additional Figures and Tables}

\begin{figure*}
  \centering
  \includegraphics{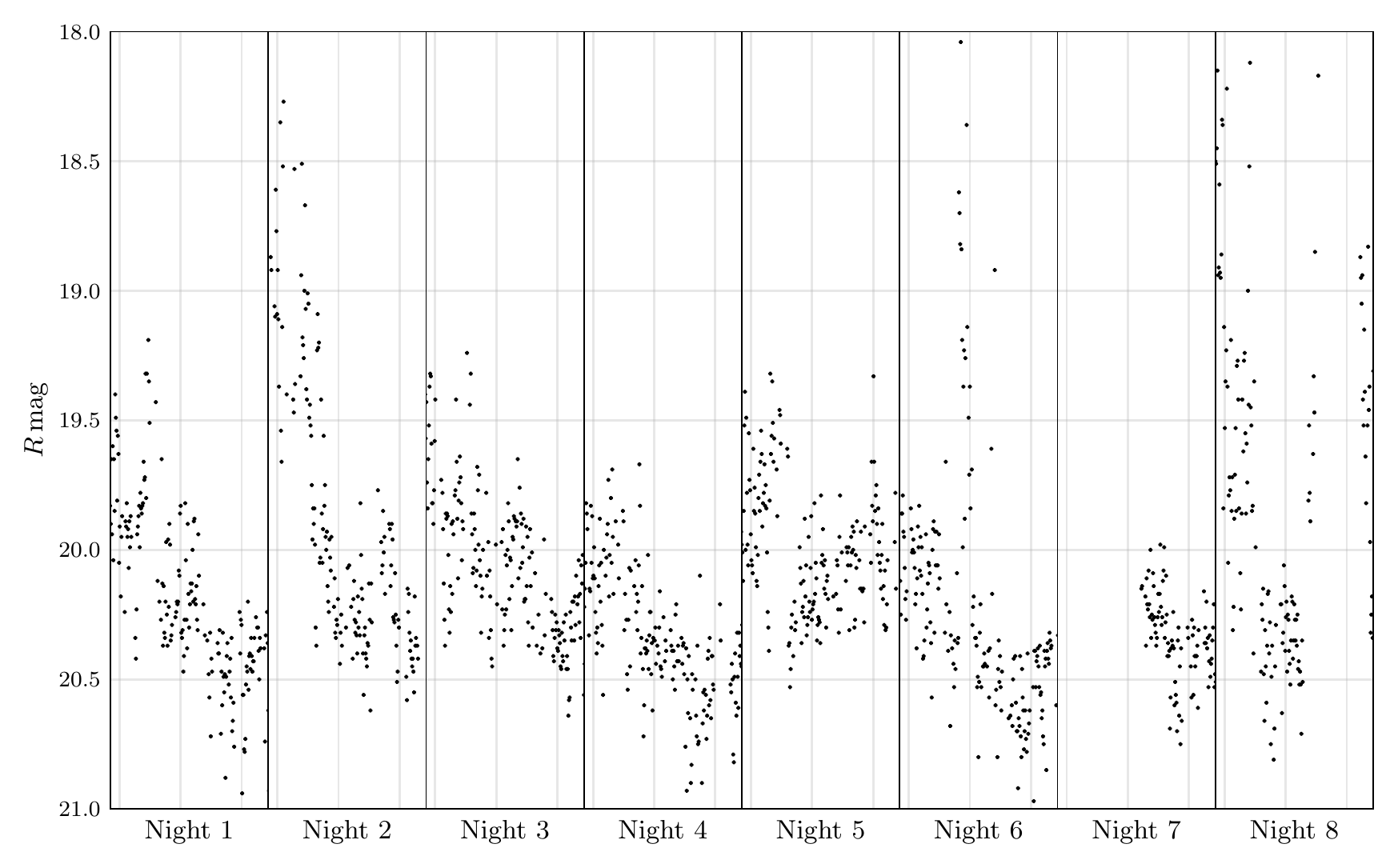}
  \caption{The limiting magnitude of the P48 images as function of time. Night 1 starts on MJD 55501.08.}
  \label{fig:PTFlimmag}
\end{figure*}

\begin{figure*}
  \centering
  \includegraphics{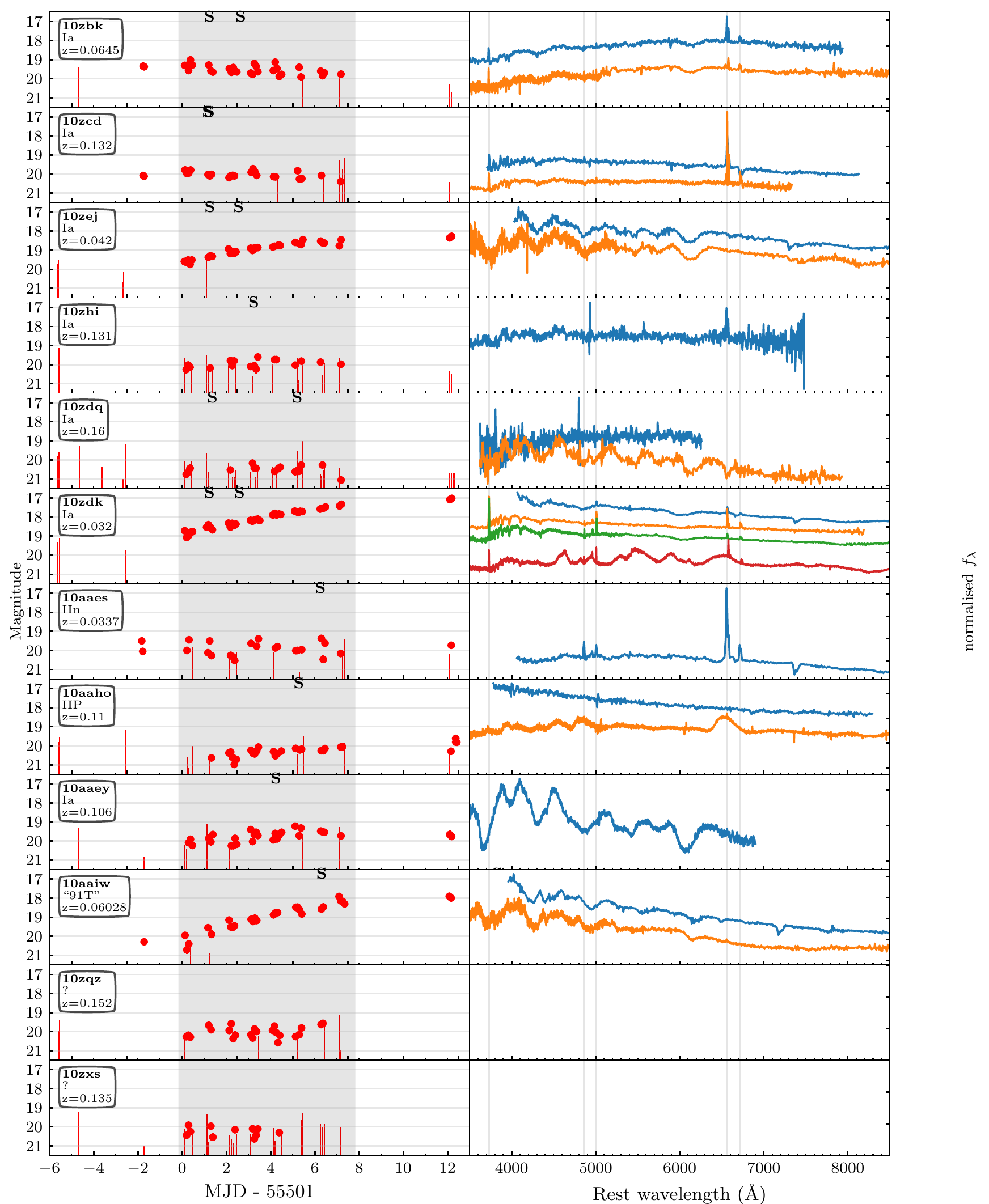}
 \caption{light curves and spectra of all supernovae found during the project. The grey shaded area indicates the duration of the Sky2Night project. The spectra are normalised to the median and offset by 1, from high to low according to time obtained and also coloured according to the time of observation from blue, yellow, green and red. The grey lines indicate common spectral lines: H\,$\alpha$, H\,$\beta$, O\,\textsc{ii}, O\,\textsc{iii} and S\,\textsc{ii}. Note that some improperly subtracted telluric lines are visible in the spectra of PTF10zhi (4940\,\AA) and for PTF10zdq (4800\,\AA). The last spectrum of PTF10zdk was taken 37 days after the start of the Sky2Night project (MJD 55538).}
  \label{fig:SNdata}
\end{figure*}

\begin{figure*}
  \centering
  \includegraphics{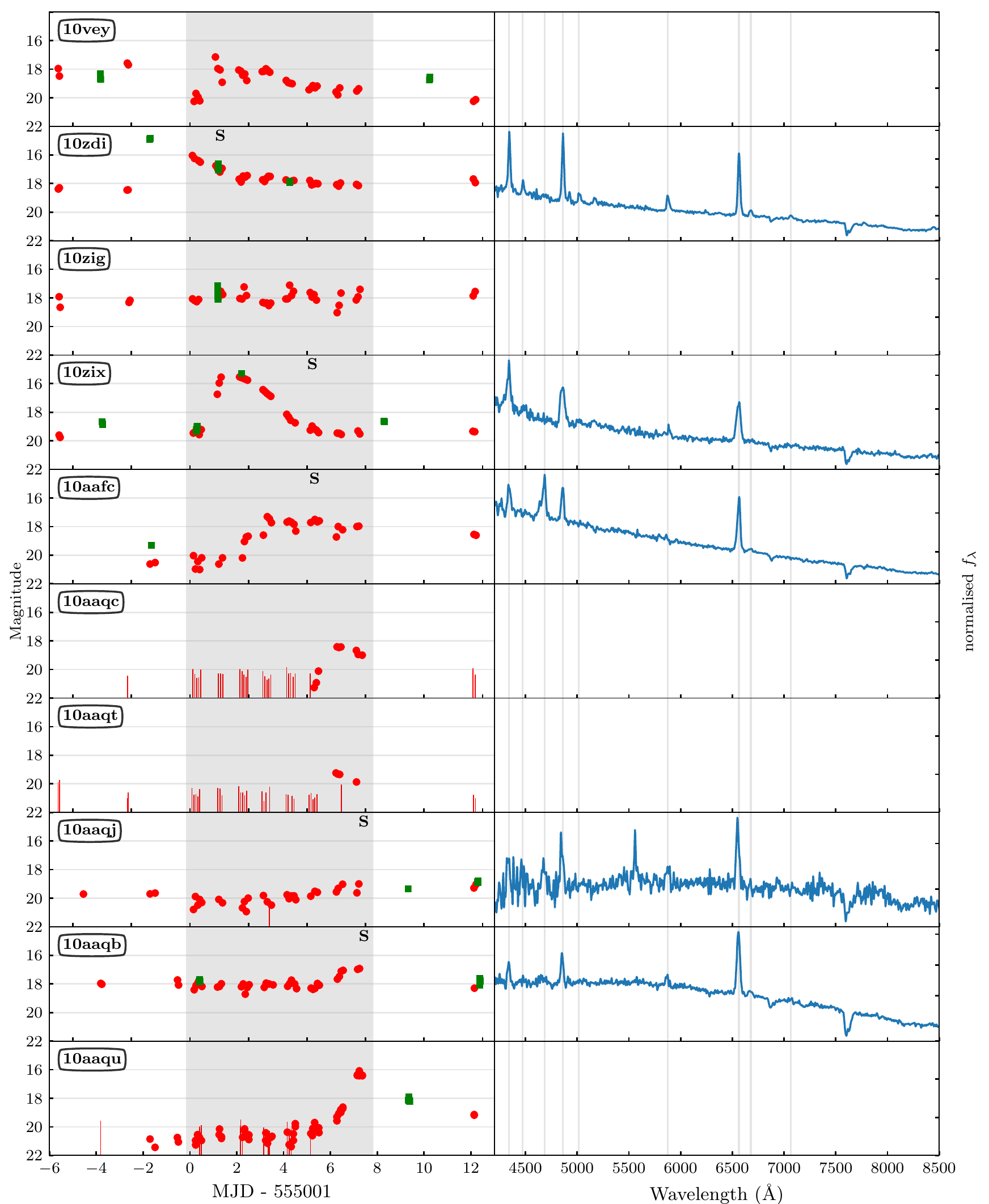}
  \caption{light curves and spectra of all cataclysmic variables found during the project. The red dots indicate PTF photometry and vertical red lines indicate upper limits ($R$ filter), green squares indicate CRTS photometry (no filter). The grey shaded area indicates the duration of the Sky2Night project. All spectra were obtained with ACAM and normalised to the mean value. Grey lines show the Balmer lines, He\,\textsc{i}, and He\,\textsc{ii} lines. The emission feature at 5577\,\AA\ in the spectrum of PTF10aaqj is caused by a telluric line.}
  \label{fig:CVdata}
\end{figure*}

\begin{figure*}
  \centering
  \includegraphics{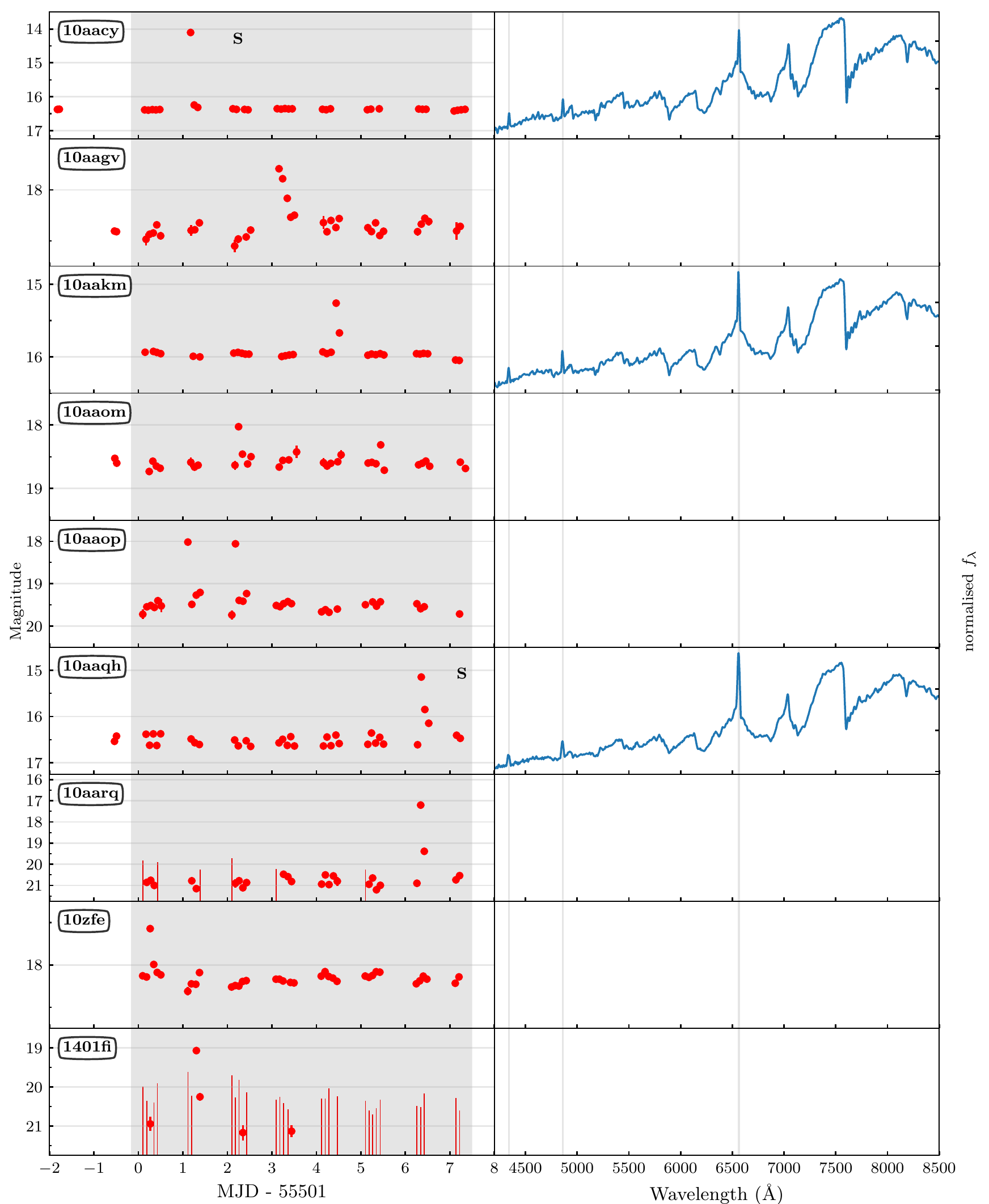}
  \caption{light curves and spectra of all stellar flares. The red dots indicate PTF photometry ($R$ filter), vertical lines indicate upper limits. The grey shaded area indicates the duration of the Sky2Night project. All spectra were obtained with ACAM and normalised to the mean value. Grey lines show the Balmer lines.}
  \label{fig:flaredata}
\end{figure*}

\begin{figure*}
  \centering
  \includegraphics{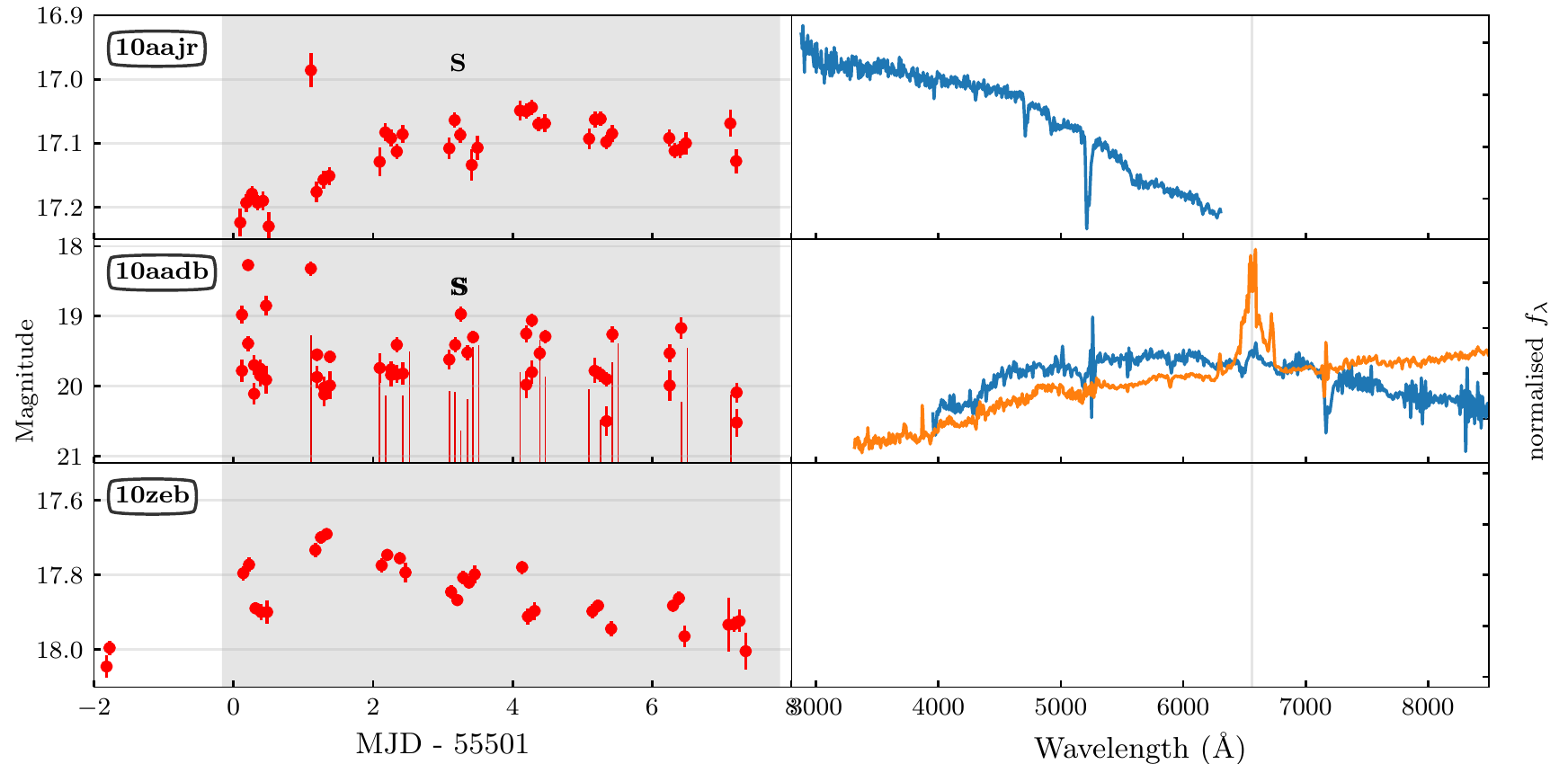}
  \caption{The light curves and spectra of flaring AGN. The red dots indicate PTF photometry ($R$ filter), vertical lines indicate upper limits. The grey shaded area indicates the duration of the Sky2Night project. The light curve for extended object PTF10aadb is made using the difference images, the other two light curves (two QSOs) are made without subtraction of the reference image. Spectra were obtained with ACAM and LRIS and normalised to the mean value. The DBSP spectrum of PTF10aadb (yellow) was taken 28 days after the start of the project while the ACAM spectrum (blue) was obtained at day 3 of the project. The grey line indicates the H$\alpha$ line.}
  \label{fig:otherdata}
\end{figure*}


\begin{table*}
\centering
\caption{The coordinates, the average extinction and brightest stars of PTF fields observed during the project.}
\label{tab:S2Nfields}
 \begin{tabular}{l|llllllll}
 FieldID  &  RA (\degree) & Dec (\degree) &     Ec. Lon. (\degree) &  Ec. Lat. (\degree) &   Gal. Lon. (\degree) & Gal. Lat. (\degree) &  $E(B-V)$ & Brightest star (mag) \\
 \hline
3430    &16.0396    &16.875    &21.2721    &9.2741     &127.306    &-45.8887    &0.045    &5.68    \\
3431    &19.604     &16.875    &24.459     &7.9678     &132.161    &-45.5128    &0.078    &3.71    \\
3432    &23.1683    &16.875    &27.6492    &6.6942     &136.916    &-44.8742    &0.089    &3.71    \\
3433    &26.7327    &16.875    &30.8459    &5.4573     &141.528    &-43.9839    &0.053    &5.21    \\
3434    &30.297     &16.875    &34.0516    &4.261      &145.96     &-42.8563    &0.061    &5.21    \\
3435    &33.8614    &16.875    &37.2688    &3.1092     &150.189    &-41.5079    &0.098    &5.89    \\
3436    &37.4257    &16.875    &40.4995    &2.0057     &154.198    &-39.9567    &0.454    &6.05    \\
3437    &40.9901    &16.875    &43.7453    &0.9542     &157.982    &-38.2211    &0.1    &5.32    \\
3438    &44.5545    &16.875    &47.0077    &-0.0417    &161.541    &-36.319    &0.224    &5.32    \\
3439    &48.1188    &16.875    &50.2875    &-0.9783    &164.88     &-34.2677    &0.129    &6.1    \\
3440    &51.6832    &16.875    &53.5855    &-1.8524    &168.01     &-32.0832    &0.123    &6.26    \\
3441    &55.2475    &16.875    &56.9018    &-2.6606    &170.943    &-29.78     &0.221    &6        \\
3442    &58.8119    &16.875    &60.2362    &-3.3998    &173.694    &-27.3716    &0.436    &5.91    \\
3531    &16.2       &19.125    &22.3031    &11.2882    &127.294    &-43.6335    &0.04    &4.77    \\
3532    &19.8       &19.125    &25.4973    &9.9791     &131.952    &-43.26     &0.054    &4.77    \\
3533    &23.4       &19.125    &28.6931    &8.7045     &136.519    &-42.6317    &0.055    &5.34    \\
3534    &27.0       &19.125    &31.8936    &7.4682     &140.956    &-41.7586    &0.059    &5.21    \\
3535    &30.6       &19.125    &35.1015    &6.2741     &145.233    &-40.6537    &0.08    &5.21    \\
3536    &34.2       &19.125    &38.3192    &5.1263     &149.324    &-39.3323    &0.136    &5.28    \\
3537    &37.8       &19.125    &41.5487    &4.0284     &153.216    &-37.8109    &0.096    &5.57    \\
3538    &41.4       &19.125    &44.7918    &2.9841     &156.9      &-36.1066    &0.085    &5.17    \\
3539    &45.0       &19.125    &48.0498    &1.9971     &160.376    &-34.2362    &0.194    &4.45    \\
3540    &48.6       &19.125    &51.3236    &1.0707     &163.647    &-32.216    &0.106    &4.45    \\
3541    &52.2       &19.125    &54.614     &0.2084     &166.722    &-30.0612    &0.169    &4.87    \\
3542    &55.8       &19.125    &57.921     &-0.5867    &169.611    &-27.7861    &0.405    &5.67    \\
3543    &59.4       &19.125    &61.2446    &-1.3117    &172.326    &-25.4034    &0.321    &5.62    \\
3631    &16.5306    &21.375    &23.4947    &13.2383    &127.489    &-41.3668    &0.041    &4.5    \\
3632    &20.2041    &21.375    &26.7269    &11.9154    &132.004    &-40.9786    &0.048    &4.77    \\
3633    &23.8775    &21.375    &29.9592    &10.6295    &136.434    &-40.3384    &0.07    &5.34    \\
3634    &27.551     &21.375    &33.1947    &9.3848     &140.743    &-39.4557    &0.07    &4.8    \\
3635    &31.2245    &21.375    &36.4363    &8.1853     &144.9      &-38.3429    &0.126    &4.8    \\
3636    &34.898     &21.375    &39.6865    &7.0349     &148.883    &-37.0148    &0.101    &5.04    \\
3637    &38.5714    &21.375    &42.9474    &5.9376     &152.679    &-35.487    &0.135    &5.47    \\
3638    &42.2449    &21.375    &46.2206    &4.897      &156.279    &-33.776    &0.342    &5.17    \\
3639    &45.9184    &21.375    &49.5076    &3.9168     &159.683    &-31.898    &0.432    &4.45    \\
3640    &49.5918    &21.375    &52.8094    &3.0004     &162.892    &-29.8688    &0.333    &4.45    \\
3641    &53.2653    &21.375    &56.1264    &2.1513     &165.913    &-27.7033    &0.185    &5.22    \\
3642    &56.9388    &21.375    &59.4589    &1.3726     &168.757    &-25.4154    &0.191    &5.43    \\
3729    &16.701     &23.625    &24.5571    &15.2451    &127.469    &-39.1113    &0.043    &4.5    \\
3730    &20.4124    &23.625    &27.7929    &13.9204    &131.812    &-38.7253    &0.065    &4.79    \\
3731    &24.1237    &23.625    &31.0272    &12.6345    &136.076    &-38.0941    &0.1    &6.24    \\
3732    &27.835     &23.625    &34.263     &11.3916    &140.229    &-37.2265    &0.119    &4.8    \\
3733    &31.5464    &23.625    &37.5033    &10.1956    &144.245    &-36.1341    &0.084    &4.8    \\
3734    &35.2577    &23.625    &40.7507    &9.0505     &148.101    &-34.8302    &0.09    &5.04    \\
3735    &38.9691    &23.625    &44.0071    &7.9601     &151.786    &-33.3295    &0.134    &5.47    \\
3736    &42.6804    &23.625    &47.2745    &6.9281     &155.289    &-31.6474    &0.207    &5.17    \\
3737    &46.3918    &23.625    &50.5542    &5.958      &158.61     &-29.799    &0.141    &5.17    \\
3738    &50.1031    &23.625    &53.847     &5.0533     &161.748    &-27.7994    &0.212    &5.46    \\
3739    &53.8144    &23.625    &57.1536    &4.2172     &164.71     &-25.6627    &0.22    &3.6    \\
3826    &17.0526    &25.875    &25.7913    &17.1841    &127.647    &-36.8431    &0.086    &4.75    \\
3827    &20.8421    &25.875    &29.0625    &15.8464    &131.867    &-36.4421    &0.119    &4.75    \\
3828    &24.6316    &25.875    &32.3304    &14.5504    &136.012    &-35.7978    &0.1    &6.25    \\
3829    &28.4211    &25.875    &35.5986    &13.3004    &140.052    &-34.9189    &0.116    &4.8    \\
3830    &32.2105    &25.875    &38.87      &12.1004    &143.961    &-33.8163    &0.063    &4.8    \\
3831    &36.0       &25.875    &42.1473    &10.9545    &147.72     &-32.503    &0.088    &5.02    \\
3832    &39.7895    &25.875    &45.4327    &9.8665     &151.315    &-30.9931    &0.153    &3.58    \\
3833    &43.5789    &25.875    &48.7279    &8.8402     &154.739    &-29.3015    &0.096    &3.58    \\
3834    &47.3684    &25.875    &52.0342    &7.8791     &157.988    &-27.443    &0.197    &5.46    \\
3835    &51.1579    &25.875    &55.3527    &6.9866     &161.063    &-25.432    &0.152    &5.64    \\
\hline
\end{tabular}
\end{table*}

\begin{table*}
\caption{Overview of the weather conditions during the observations. The time inbetween astronomical twilight was approximately 12.3\,h for both PTF and WHT.}
\label{tab:weather}
\begin{tabular}{l l|llllllll}
 & &  Night 1 & Night 2 & Night 3 & Night 4 & Night 5 & Night 6 & Night 7 & Night 8 \\
\hline
\hline
\multirow{3}{*}{PTF}  & time lost (h) & 0.6 & 3.7 & 0.1 & 0.2 & 0.7 & 0.4 & 3.7 & 6.0 \\
                      & seeing (\arcsec) & 2.5-3.5 & 3.0 & 3.0 & 2.5 & 2.0-3.5 & 2.0 & 3.0 & 2.5-3.5 \\
                      & cloud conditions & good & good & good & ok & ok & ok & bad/ok & bad \\
\hline
\multirow{3}{*}{WHT} & time lost (h) & 0 & 3.0 & 10.3 & 0 & 0 & 0 & 6.7 & \\
                     & seeing (\arcsec) & 1.5-2.5 & 1.5-3.0 & 2.5 & 2.5 & 1.5-2.5 & 1.0 & 2-3 & \\
                     & cloud conditions & good & ok-bad & bad & good & good & good & good &  \\
\end{tabular}
\end{table*}

\begin{table*}
\caption{Number of observations of each field. Night 1 starts at MJD 55501.08.}
\label{tab:nobs}
\begin{tabular}{l|llllllll|l}
FieldID & Night 1 & Night 2 & Night 3 & Night 4 & Night 5 & Night 6 & Night 7 & Night 8 & Total \\
\hline
3430 & 4 & 3 & 4 & 4 & 3 & 4 & 2 & 4 & 28 \\
3431 & 4 & 3 & 5 & 5 & 3 & 4 & 2 & 4 & 30 \\
3432 & 4 & 3 & 5 & 5 & 3 & 4 & 3 & 4 & 31 \\
3433 & 5 & 3 & 4 & 5 & 3 & 3 & 3 & 4 & 30 \\
3434 & 5 & 3 & 5 & 5 & 3 & 4 & 3 & 4 & 32 \\
3435 & 5 & 3 & 5 & 5 & 4 & 5 & 3 & 4 & 34 \\
3436 & 5 & 4 & 5 & 5 & 3 & 4 & 3 & 3 & 32 \\
3437 & 5 & 3 & 5 & 4 & 3 & 5 & 3 & 3 & 31 \\
3438 & 5 & 3 & 4 & 5 & 3 & 5 & 3 & 3 & 31 \\
3439 & 5 & 3 & 5 & 5 & 4 & 5 & 3 & 3 & 33 \\
3440 & 5 & 3 & 5 & 5 & 4 & 5 & 4 & 3 & 34 \\
3441 & 4 & 3 & 5 & 5 & 5 & 5 & 4 & 3 & 34 \\
3442 & 5 & 3 & 5 & 4 & 5 & 5 & 4 & 3 & 34 \\
3531 & 5 & 5 & 4 & 5 & 5 & 4 & 2 & 3 & 33 \\
3532 & 5 & 2 & 4 & 5 & 5 & 4 & 3 & 3 & 31 \\
3533 & 4 & 3 & 4 & 5 & 5 & 4 & 3 & 3 & 31 \\
3534 & 5 & 4 & 4 & 5 & 5 & 4 & 3 & 3 & 33 \\
3535 & 5 & 4 & 5 & 5 & 5 & 5 & 3 & 4 & 36 \\
3536 & 4 & 4 & 4 & 5 & 5 & 5 & 3 & 4 & 34 \\
3537 & 5 & 4 & 5 & 5 & 5 & 5 & 3 & 4 & 36 \\
3538 & 4 & 4 & 5 & 6 & 6 & 5 & 3 & 3 & 36 \\
3539 & 5 & 3 & 5 & 5 & 4 & 5 & 4 & 3 & 34 \\
3540 & 5 & 3 & 5 & 5 & 5 & 5 & 4 & 3 & 35 \\
3541 & 5 & 3 & 5 & 5 & 5 & 5 & 4 & 3 & 35 \\
3542 & 4 & 3 & 5 & 5 & 5 & 6 & 4 & 3 & 35 \\
3543 & 4 & 3 & 5 & 5 & 5 & 5 & 4 & 2 & 33 \\
3631 & 5 & 3 & 5 & 5 & 4 & 4 & 2 & 2 & 30 \\
3632 & 5 & 4 & 5 & 5 & 4 & 4 & 3 & 3 & 33 \\
3633 & 5 & 4 & 5 & 5 & 3 & 5 & 3 & 2 & 32 \\
3634 & 5 & 2 & 5 & 5 & 5 & 5 & 3 & 2 & 32 \\
3635 & 5 & 4 & 5 & 5 & 4 & 5 & 3 & 2 & 33 \\
3636 & 6 & 3 & 5 & 6 & 5 & 5 & 3 & 1 & 34 \\
3637 & 5 & 4 & 5 & 5 & 5 & 5 & 3 & 3 & 35 \\
3638 & 5 & 2 & 5 & 6 & 5 & 5 & 4 & 3 & 35 \\
3639 & 5 & 3 & 5 & 5 & 5 & 4 & 4 & 2 & 33 \\
3640 & 5 & 3 & 5 & 5 & 5 & 5 & 4 & 2 & 34 \\
3641 & 5 & 3 & 5 & 4 & 5 & 5 & 4 & 2 & 33 \\
3642 & 5 & 3 & 5 & 5 & 5 & 5 & 4 & 2 & 34 \\
3729 & 5 & 3 & 5 & 5 & 5 & 4 & 3 & 2 & 32 \\
3730 & 4 & 4 & 5 & 5 & 4 & 5 & 3 & 2 & 32 \\
3731 & 5 & 4 & 5 & 4 & 5 & 5 & 4 & 1 & 33 \\
3732 & 6 & 4 & 5 & 5 & 4 & 5 & 4 & 2 & 35 \\
3733 & 6 & 4 & 5 & 5 & 5 & 5 & 4 & 2 & 36 \\
3734 & 5 & 4 & 5 & 6 & 5 & 5 & 4 & 2 & 36 \\
3735 & 5 & 4 & 6 & 6 & 5 & 5 & 4 & 1 & 36 \\
3736 & 5 & 4 & 5 & 5 & 5 & 6 & 4 & 2 & 36 \\
3737 & 5 & 3 & 5 & 5 & 5 & 5 & 4 & 2 & 34 \\
3738 & 5 & 3 & 5 & 5 & 5 & 5 & 4 & 2 & 34 \\
3739 & 5 & 3 & 5 & 5 & 5 & 5 & 4 & 2 & 34 \\
3826 & 5 & 4 & 5 & 5 & 4 & 5 & 3 & 2 & 33 \\
3827 & 5 & 4 & 5 & 4 & 5 & 5 & 3 & 2 & 33 \\
3828 & 5 & 4 & 5 & 5 & 4 & 5 & 3 & 2 & 33 \\
3829 & 6 & 4 & 5 & 5 & 5 & 5 & 3 & 2 & 35 \\
3830 & 6 & 4 & 5 & 5 & 4 & 5 & 3 & 1 & 33 \\
3831 & 5 & 4 & 6 & 6 & 5 & 6 & 3 & 2 & 37 \\
3832 & 5 & 4 & 4 & 6 & 5 & 6 & 4 & 2 & 36 \\
3833 & 5 & 3 & 5 & 5 & 5 & 6 & 4 & 2 & 35 \\
3834 & 5 & 3 & 4 & 5 & 5 & 5 & 4 & 2 & 33 \\
3835 & 5 & 3 & 5 & 5 & 5 & 5 & 4 & 2 & 34 \\
\hline
median & 5 &  3 & 5 & 5 & 5 & 5 & 3 &  2 & 34 \\
\hline
total & 290 & 200 & 287 & 296 & 266 & 285 & 199 & 151 & 1974 \\
\end{tabular}
\end{table*}

\begin{table}
\centering
\caption{Properties of the supernovae discovered during the Sky2Night project.}
\label{tab:SNstats}
\begin{tabular}{llll}
Name          & Type       & Redshift      & peaktime       \\
PTF ...       &            & (z)           & (MJD-55501)    \\
\hline
10zbk    & Ia      & $0.0645(5)$  & $-9 \pm 3$     \\
10zcd    & Ia      & $0.132(2)$   & $-1.9 \pm 1.8$ \\
10zej    & Ia ``91bg'':  & $0.048(6)$   & $11.5 \pm 0.2$ \\
10zhi    & Ia      & $0.128(5)$   & $5.5 \pm 1.4$  \\
10zdq    & Ia      & $0.161(4)$   & $0.3 \pm 3.0$  \\
10zdk    & Ia      & $0.033(1)$   & $14.2 \pm 0.1$ \\
10aaes   & IIn     & $0.0337(1)$  & $<-80$         \\
10aaho   & IIP     & $0.108(4)$   & $6.5 \pm 0.4$  \\
10aaey   & Ia      & $0.107(2)$   & $8.4 \pm 1.0$  \\
10aaiw   & Ia ``99T''  & $0.06028(2)$ & $16.3 \pm 0.3$ \\
10zxs    & ?       & $0.135(34)$ & $ $ \\
10zqz    & ?       & $0.152(38)$ & $ $ \\
\hline
\end{tabular}
\end{table}

\begin{table}
\centering
\caption{Properties of the CVs found during the Sky2Night project. The quiescence magnitude is given in $R$ if it is detected in PTF images. If the counterpart is too faint, the PanSTARRS $r$ magnitude is given.}
\label{tab:CV}
\begin{tabular}{lllll}
Name          & type & quiescence & $\Delta r$             \\
PTF ...       &         & (mag)   & (mag)           \\
\hline
10vey    & U Gem  & $R=20.5$   & 3.4  \\
10zdi    & U Gem  & $R=18.4$   & 2.4  \\
10zig    & SU UMa:/WZ Sge:  & $r=21.6$ & 5.5  \\
10zix    & U Gem  & $R=19.61$   & 3.9  \\
10aafc    & U Gem  & $R=20.17$   & 2.5  \\
10aaqc    & U Gem  & $r=22.0$   & 3.5  \\
10aaqt    & U Gem  & $r=23.0$   & 4.5   \\
10aaqj    & AM Her:  & $R=20.5$ & 1.0   \\
10aaqb    & U Gem  & $R=18.00$   & 1.2  \\
10aaqu    & U Gem  & $R=20.34$   & 4.3  \\
\hline
\end{tabular}
\end{table}

\begin{table}
\centering
\caption{Properties of the flaring M-stars discovered during the Sky2Night project. PTF10aaop showed two outbursts in different nights.}
\label{tab:dMe}
\begin{tabular}{llllll}
Name          & Sp type & quiescence $R$ &  $\Delta R$      & time scale & $\mathrm{\log E_R} $      \\
PTF ...       &         & (mag)   &    (mag)  & (h)    & $(\mathrm{erg s^{-1}})$\\
\hline
10aacy    & M4  & 16.4   & 2.3  & 0.5(1) & 32.0 \\
10aagv    & M5  & 18.3   & 0.6  & 4.6(4) & 32.0 \\
10aakm    & M4  & 15.9   & 0.6  & 1.6(1) & 32.0 \\
10aaom    & M5  & 18.7   & 0.7  & 1.3(2) & 31.4 \\
10aaop    & M7  & 19.5   & 1.5/1.5  & <0.3/<0.6 & 30.0/30.4 \\
10aaqh    & M5  & 16.5   & 1.3  & 1.9(1) & 32.1 \\
10aarq    & M6  & 20.7   & 3.5  & 0.8(1) & 31.6 \\
10zfe     & M4  & 18.2   & 0.6  & 1.4(1) & 32.0 \\
1401fi    & M4  & 21.3   & 2.2  & 1.2(2) & 34.0 \\
\hline
\end{tabular}
\end{table}


\bsp    
\label{lastpage}
\end{document}